 \documentclass[preprint]{aastex}

\begin{document}

\title{AGES: The AGN and Galaxy Evolution Survey\footnote{Observations reported here were obtained 
      at the MMT Observatory, a joint facility of the Smithsonian Institution and the University of Arizona. } }
\shorttitle{AGES}
\shortauthors{C.S. Kochanek et al.}
   \author{C.S. Kochanek\altaffilmark{1,2},
           D.J. Eisenstein\altaffilmark{3},
           R.J.  Cool\altaffilmark{4},
           N. Caldwell\altaffilmark{3},
           R.J. Assef\altaffilmark{5},
           B.T. Jannuzi\altaffilmark{6},
           C. Jones\altaffilmark{3},
           S.S. Murray\altaffilmark{3,7}, 
           W.R. Forman\altaffilmark{3},
           A. Dey\altaffilmark{6},
           M.J.I. Brown\altaffilmark{8},
           P. Eisenhardt\altaffilmark{5},
           A.H. Gonzalez\altaffilmark{9}
           P. Green\altaffilmark{3},
           D. Stern\altaffilmark{5}
         }

   \altaffiltext{1}{Department of Astronomy, Ohio State University, 140 West
              18th Avenue, Columbus, OH 43210}
   \altaffiltext{2}{Center for Cosmology and Astroparticle Physics, Ohio State University, 191 W. Woodruff Avenue,
              Columbus, OH 43210}
   \altaffiltext{3}{Harvard-Smithsonian Center for Astrophysics, 60 Garden Street, Cambridge MA 02138}
   \altaffiltext{4}{Hubble and Carnegie/Princeton Fellow,
             Department of Astrophysical Sciences, Princeton University, Peyton Hall, Princeton NJ 08544 and
             The Observatories of the Carnegie Institution of Washington, 813 Santa Barbara Street, Pasadena, CA 91101}
   \altaffiltext{5}{Jet Propulsion Laboratory, California Institute of Technology, 4800 Oak Grove Drive, 
             Pasadena, CA 91109}
   \altaffiltext{6}{NOAO, 950 North Cherry Avenue, Tucson, AZ, 85719}
   \altaffiltext{7}{Department of Physics and Astronomy, Johns Hopkins University, Baltimore, MD, 21218}
   \altaffiltext{8}{School of Physics, Monash University, Clayton, Victoria 3800, Australia}
   \altaffiltext{9}{Department of Astronomy, Bryant Space Science Center, University of Florida, Gainesville, FL 32611, USA }

\begin{abstract}
The AGN and Galaxy Evolution Survey (AGES) is a redshift survey covering, in its standard
fields,  $7.7$~deg$^2$
of the Bo\"otes field of the NOAO Deep Wide-Field Survey (NDWFS).  The
final sample consists of 23745 redshifts.  There are well-defined galaxy samples in
ten bands (the B$_W$, R, I, J, K, IRAC $3.6$, $4.5$, $5.8$ and $8.0\mu$m and 
MIPS $24\mu$m bands) to a limiting magnitude of $\hbox{I}<20$~mag for
spectroscopy.  For these galaxies, we obtained 18163 redshifts from a sample of 35200 galaxies, where
random sparse sampling was used to define statistically complete sub-samples in all
ten photometric bands. The median 
galaxy redshift is $0.31$, and 90\% of the redshifts are in the range $0.085 < z < 0.66$.
AGN were selected as radio, X-ray, IRAC mid-IR and MIPS $24\mu$m sources to fainter
limiting magnitudes ($\hbox{I}<22.5$~mag for point sources).  Redshifts were obtained for
4764 quasars and galaxies with AGN signatures, with 2926, 1718, 605, 119 and 13 above redshifts of
$0.5$, $1$, $2$, $3$ and $4$, respectively.  We detail all the AGES selection procedures
and present the complete spectroscopic redshift catalogs, spectra, and spectral energy
distribution decompositions.  Photometric redshift estimates are for all 
sources in the AGES samples.  
\end{abstract}


\def\imag{\hbox{I}}
\def\jmag{\hbox{J}}
\def\kmag{\hbox{K}}
\def\rmag{\hbox{R}}
\def\bmag{\hbox{B}_W}
\def\zmag{\hbox{z'}}
\section{Introduction}

Surveys are a critical tool for understanding the evolution of
galaxies and AGN.  Because their properties are diverse and
changing, we utilize large statistical samples of galaxies to measure
the distribution of their properties and to trace the evolution 
of these distributions.
Achieving a suitably high level of detail requires a combination
of multiwavelength imaging and spectroscopy.  Without a redshift
estimate, one cannot infer luminosity, color, or environmental
density, all quantities known to be of central importance in the
behavior of galaxies and AGN.  While photometric redshifts can
be used in the absence of spectroscopy, they have more systematic
uncertainties (e.g. \citealt{Hildebrandt2010}) and have difficulty
with AGN (e.g. \citealt{Brodwin2006}, \citealt{Rowan2008}, \citealt{Assef2010}). Moreover, the story of galaxy evolution
involves many wavelengths of light: UV and far-IR for young stars,
optical and near-IR for older stars, and X-ray, radio and mid-IR for nuclear
activity.

Decades of surveys have quantified the luminosity, color, surface
brightness, star formation, and nuclear activity of low redshift
galaxies and correlated these properties with environment.  The
combination of the CfA Redshift Survey (\citealt{deLapparent1986}) and the 
Palomar Sky Survey (POSS-II, \citealt{Reid1991})
defined the state of the art for local galaxies in the 1980's.  This
was followed by the Las Campanas Redshift Survey and its drift-scan
CCD imaging (\citealt{Shectman1996}), which expanded our view to larger scales.  Most recently,
the 2 Degree Field Galaxy Redshift Survey (\citealt{Colless2001}) and the Sloan Digital Sky
Survey (SDSS \citealt{York2000}) brought the scale of spectroscopy to the million-galaxy
level.  Wide area digital imaging from the 2 Micron All-Sky Survey (2MASS, \citealt{Skrutskie2006}), SDSS,
GALEX (\citealt{Martin2005}), the Spitzer Space Telescope (\citealt{Werner2004}), 
and ROSAT (e.g. \citealt{Voges1999}) have been combined with this spectroscopy to build a detailed 
characterization of nearby galaxies.

Surveys at higher redshift require much deeper imaging and fainter
spectroscopy.  Projects such as GOODS (\citealt{Giavalisco2004}), DEEP-2 (e.g. \citealt{Faber2007}), 
VVDS (\citealt{LeFevre2005}), 
and COSMOS (\citealt{Scoville2007}) now
aim to survey cosmologically interesting volumes at redshifts of
order unity and above.  Importantly, there is substantial evolution
in galaxy properties.  Since $z\approx1$, the star formation rate
per unit comoving volume has dropped by a factor of 10 (e.g. \citealt{Hopkins2006}), the frequency
of luminous quasars (e.g. \citealt{Croom2004}, \citealt{Richards2005},
\citealt{Richards2006}) and ultra-luminous infrared galaxies have decreased by a
factor of 100 (e.g. \citealt{Cowie2004}, \citealt{Lefloch2005}), the mass of galaxies on the red sequence has
roughly doubled (e.g. \citealt{Bell2003}, \citealt{Faber2007}), and the typical site for star formation has moved
to less massive galaxies (e.g. \citealt{Cowie1996}).  At redshifts above unity, further evolution
is clear, with galaxies getting notably smaller (e.g. \citealt{Daddi2005}, \citealt{vandokkum2010}), possibly with
changing correlations of star formation with environment (e.g. \citealt{Scodeggio2009}, \citealt{Cooper2010}).

Mapping galaxy properties at intermediate redshift ($z\sim0.5$) and
the demographics of active galactic nuclei (AGN) at any redshift
requires wider fields than these deep surveys can provide.  We
designed the AGN and Galaxy Evolution Survey (AGES) to address these
questions, combining spectroscopy from the Hectospec instrument on
the MMT (\citealt{Fabricant1998},
\citealt{Roll1998}, \citealt{Fabricant2005}) with superb multi-wavelength imaging in the NOAO Deep 
Wide-Field Survey (NDWFS) Bo\"otes field.  This field contains deep imaging at
optical and near-IR bands (\citealt{Jannuzi1999}, \citealt{Elston2006}) as well as full-field coverage from Spitzer
IRAC (\citealt{Eisenhardt2004} and later \citealt{Ashby2009}) and  MIPS (\citealt{Soifer2004}), 
Chandra (\citealt{Murray2005}, \citealt{Kenter2005}, \citealt{Brand2006}), GALEX (\citealt{Martin2005}), and radio (\citealt{Becker1995}, 
\citealt{deVries2002}) 
facilities.  As we detail below, AGES provides
spectroscopic redshifts for 18163 galaxies to $I=20$ and 4764 AGN
to $I=22.5$ in the 9 square degree field.

AGES sought to exploit the diversity of available imaging data by 
a multi-faceted targeting strategy.  AGN were selected by optical
color, IRAC color, and MIPS, X-ray, or radio detection.  The result is a
large and broad sample of AGN, including nearly 200 per square degree
at $z>1$ and nearly 400 per square degree in total.  The galaxy
sample required sparse sampling to reach $I=20$, but we tuned the
sparse sampling algorithm to ensure full sampling of brighter 
magnitude thresholds in ten different bands from the UV to the mid-IR.
The AGES observational strategy returned to the field many times with
rolling acceptance, with the result of very high completeness in
the statistical samples.  This also means that fiber collisions are an
unimportant problem even in high density regions. AGES is well tuned for the study of galaxy
properties at $0.2<z<0.6$ and AGN properties out to $z=5$ over a
cosmologically sized volume with pan-chromatic spectral energy
distributions.

This paper presents the AGES data set and data release.  Section 2
describes survey design and target selection.  Section 3 discusses
the observations themselves.  Section 4 describes the data reduction
procedures.  Section 5 summarizes the resulting samples and Section
6 introduces the files of the data release.  We conclude in Section 7.
All magnitudes and fluxes are in the system used by the parent 
survey.  These are Vega magnitudes for the B$_w$, R, I, J, K, K$_s$
and IRAC bands, AB magnitudes for the FUV, NUV and $z'$ bands, mJy for  
MIPS $24\mu$m band and the radio observations, and $0.5$-$7$~keV
counts for the X-ray observations.

\section{Survey Design}
\label{sec:design}

We selected targets at all wavelengths from radio through X-ray to take
full advantage of the available imaging data.  We start with the optical
data of the NDWFS itself in the B$_W$, R and I bands (\citealt{Jannuzi1999}), since our
ability to measure spectra is limited by the optical flux.  We used
the zBo\"otes (\citealt{Cool2007}) data to help target high-redshift
quasars. In the 
near-infrared we used the K-band data of the NDWFS and the J/K$_s$-band
data of the FLAMEX survey (\citealt{Elston2006}).  In the mid-infrared
we used the 3.6, 4.5, 5.8 and 8.0$\mu$m data from the IRAC Shallow Survey
(\citealt{Eisenhardt2004}).  We used the 24$\mu$m data from \cite{Soifer2004}.
In the radio we used the FIRST survey (\citealt{Becker1995}) and the deeper
1.4~GHz WSRT catalog of \cite{deVries2002}.  Going to shorter
wavelengths we used data from GALEX (\citealt{Martin2005}) in the UV and the
Chandra XBo\"otes survey (\citealt{Murray2005}, \citealt{Kenter2005}, \citealt{Brand2006})
at X-ray wavelengths.

Our general approach was to produce well-defined, magnitude limited 
samples of galaxies at all the available wavelengths from 24$\mu$m 
through the GALEX FUV bands, and to target AGN using a broad range
of selection methods to a somewhat deeper optical flux limit.  The
galaxy samples were generally designed to be complete to an intermediate
magnitude limit and then randomly sparse sampled from this intermediate
limit to the overall magnitude limit.  The sample definitions changed several
times, with the largest change being a shift from preliminary
photometric catalogs and R-band magnitude limits to revised 
catalogs and I-band magnitude limits between the 2004 and 2005 
observing seasons.  

Sparse sampling was an essential component of our strategy for studying
galaxies because it was the only means of covering such a wide area 
to the desired depth ($\hbox{I}<20$~mag) in reasonable time.  To begin with,
sparse sampling over a wider area produces samples with less cosmic
variance than complete samples in smaller areas surveying smaller 
cosmological volumes.  It also allows us to sample galaxies with
very different properties with a well-defined statistical approach.
In particular, by making the survey complete for bright objects and
sparse sampling across many different photometric bands for fainter
objects, we have a final sample with less shot noise for bright 
objects and the relatively rarer objects with extreme colors in any
band.  While the approach is more complex than previous redshift surveys, it 
is nonetheless straight forward to construct a complete statistical 
sample by weighting each object by the inverse of its sampling fraction  
We provide explicit instructions on the appropriate procedures in
\S\ref{sec:release}.  

In \S2.1, 2.2 and 2.3 we define nomenclature, outline our approach
to random sparse sampling, and define the standard AGES sub-fields.  In
\S2.4, 2.5, and 2.6 we outline the sample definitions used in 2004,
2005 and 2006/2007.  Samples defined in the previous years continued
to be observed at the same priorities in the later years.  The biggest
differences were between 2004 and the later years, where we (1) shifted 
from an R-band limited sample in 2004 to an I-band limited sample, (2) 
shifted from preliminary NDWFS and IRAC Shallow Survey catalogs 
to later versions, and (3) added mid-IR quasar selection.  The main
differences from 2005 to 2006/2007 were to add sub-samples further
exploring mid-IR quasar selection and to include the zBo\"otes 
data as a tool for AGN selection.  There are many other small differences that are detailed
in each subsection. We provide the selection codes for all seasons, so that it
is clear why any source may have been targeted, but in the data tables
we only provide the photometric information for the updated catalogs
used from 2005 onwards.   The photometric data are only a limited 
representation of the underlying catalogs -- the original survey 
catalogs should be consulted to obtain the complete data.
In each season we targeted at low priority 
some objects that were not part of the primary AGES project  
as experiments from collaborators 
We very briefly outline these experiments 
and include their selection codes but provide no details. 

\begin{figure}[p]
\centerline{\includegraphics[width=5.0in]{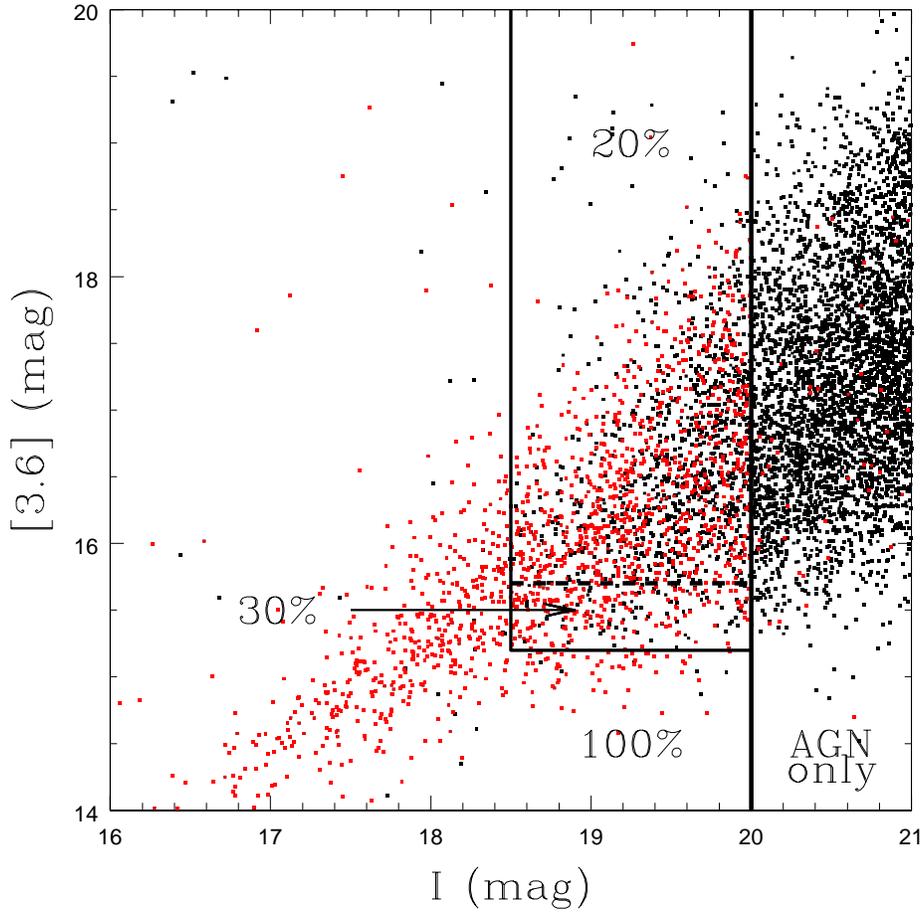}}
\caption{  The distribution of galaxies in I band and IRAC [3.6] magnitudes.  The points are
   a randomly selected 10\% of the sources in the main survey area, where red points have
   measured redshifts.  The boundaries indicate the survey sampling regions and the
   sparse sampling rates.  The actual fractions of redshift measurements in these
   regions are much higher than the nominal 20\% or 30\% sampling rates
   because of the different color weightings
   of the various bands and the observations of lower priority sources with unused fibers.   
   AGN were targeted to fainter magnitudes, leading to the redshift measurements with
   $I>20$~mag.
   }
\label{fig:select}
\end{figure}

\subsection{Common Definitions}

We will refer to the survey bands as X, FUV, NUV, B$_W$, R, I, z', J, K, [3.6]$_i$, [4.5]$_i$, [5.8]$_i$, [8.0]$_i$,
$[24]$, FIRST, and WSRT.  For the X-ray sources, X is the X-ray counts from the
XBo\"otes survey.  For the ultraviolet sources, FUV and NUV are the two GALEX filters.
For the optical and near-IR, B$_W$, R, I, z', J and K are the SExtractor Kron-like (mag$\_$auto) magnitudes.
The K refers to both the NDWFS K/K$_s$ and the FLAMEX K$_s$ data.  The final NDWFS
magnitudes are from DR3 (http://www.noao.edu/noao/noaodeep/DR3/dr3-data.html). For the IRAC 
data, [3.6]$_i$--[8.0]$_i$ are the 3.6, 4.5, 5.8 and 8.0$\mu$m SExtractor Kron-like magnitudes where
the subscript $i=[3.6]$--$[8.0]$ defines the IRAC band used to define the extraction apertures.  
The 24$\mu$m flux $[24]$ is the DAOPHOT PSF-fit flux of the source.  FIRST and WSRT
both measured 1.4~GHz radio continuum fluxes.  Where we are using the flux in a
fixed aperture, we add the aperture diameter to the magnitude, so [3.6]$_{[3.6]} (6\farcs0)$
represents the IRAC 3.6$\mu$m flux in a 6\farcs0 diameter aperture whose position was 
determined from the 3.6$\mu$m image.   These IRAC aperture magnitudes are, however,
corrected for the extension of the IRAC point spread function beyond the aperture.

For the optical data we homogenized several aperture magnitudes for seeing variations.  
For each field, we took stars in the magnitude range $19 < I < 20$ and computed the mean 
differences between the Kron-like I magnitude, presumed to be seeing independent,
and the $1\farcs0$, $3\farcs0$ and $6\farcs0$ aperture magnitudes.  
These differences, which show the expected pattern of being significant for
the $1\farcs0$ apertures and negligible of the $6\farcs0$ apertures,
were then applied to these three aperture magnitudes for the B$_W$, R and
I bands.

Point sources ($\hbox{pntsrc}=1$) were defined based on the SExtractor stellarity indices of the sources
in the optical (B$_W$, R, I and z'-bands).  For each target we assigned a code (bgood, rgood, igood, zgood$=1$) 
for whether the data in each band on that target was acceptable.  
We flagged objects as either quasar candidates ($\hbox{qso}=1$), galaxies 
($\hbox{galaxy}=1$), or in the later seasons AGN-galaxy ($\hbox{agngalaxy}=1$) targets.  
Quasar candidates are point sources brighter than the (optical) magnitude limit for targeting 
quasars, galaxies are extended sources brighter than the magnitude limit for targeting galaxies, 
and ``AGN-galaxies'' are extended AGN candidates brighter than the magnitude limit for 
targeting quasars but fainter than that for galaxies.  In general, our sources are much brighter
than the NDWFS survey limits, so there are few issues with star/galaxy separation.

\subsection{Sparse Sampling Codes}

For many of our samples we observed all targets to an intermediate magnitude limit and then
randomly sparse sampled the sources between the intermediate limit and the overall flux limit
of the sample.  All sources were assigned a random integer code $0\leq \hbox{rcode}<20$ dividing the sources
into 20 random sub-samples each containing 5\% of the targets.  Once assigned to a source, these codes
were preserved in all future samples.  The sparse sampling fraction was then determined by
the limit on the $\hbox{rcode}$ used to define the sample.  For objects that were not included
in any of the primary samples, we assigned lower $\hbox{rcode}$ values higher observing 
priorities than higher $\hbox{rcode}$ values.  This increases the completeness of the observations 
for the lower $\hbox{rcode}$ targets, so that any decision to move to a higher sparse sampling 
fraction than used initially requires observations of fewer targets while ensuring that the
fibers stay filled.

Fig.~\ref{fig:select} illustrates how this works for the IRAC [3.6] sample in 2006/2007 (see below).
  All galaxies
   must have $\hbox{I}<20$~mag, so no fainter objects were targeted unless they
   appeared in one of the AGN samples.  At I band, all galaxies were targeted if they
   were brighter than $\hbox{I}<18.5$~mag, and a randomly selected 20\% ($\hbox{rcode}\leq 3$)
   were targeted
   between $\hbox{18.5} < \hbox{I} < 20$.  For the IRAC [3.6] band, galaxies
   were all targeted if brighter than $[3.6]<15.2$~mag, and a randomly selected 30\% ($\hbox{rcode}\leq 5$)
   were targeted between $15.2<[3.6]<15.7$, but still subject to the $\hbox{I}<20$~mag
   limit.  Combining these criterion, all sources to the left or below the heavy
   solid line, 30\% of the sources in the box $18.5 < \hbox{I} < 20$ and
   $15.2 < [3.6] < 15.7$, and 20\% of the sources with $18.5 < \hbox{I} < 20$ and
   $[3.6]>15.7$~mag  were targeted by these criterion.  Between the different
   color weightings of the bands and the filling of fibers that could not be
   allocated to the primary samples, the actual fractions of sources with redshift measurements
   in the sparse sampling regions are much higher than 20\% or 30\%.

\begin{figure}[p]
\centerline{\includegraphics[width=5.0in]{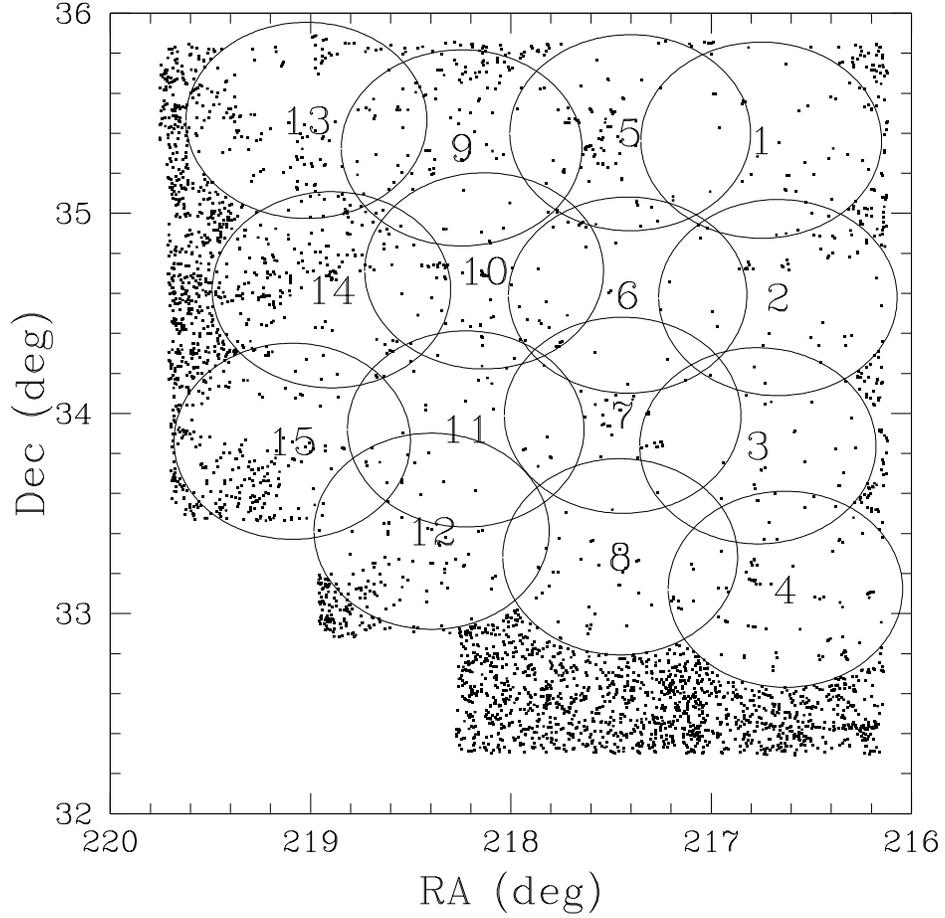}}
\caption{ The 15 standard sub-fields listed in Table~\ref{tab:fields}.  The points are the
  2006/2007 primary galaxy sample targets (gcode06 neither zero nor 2048)  
  lacking redshifts. Note the high completeness in the sub-fields and the significantly 
  lower completeness elsewhere in the Bo\"otes field.  Some redshifts were obtained outside
  the standard fields because of the shifting centers of the individual Hectospec pointings.
   }
\label{fig:fields}
\end{figure}

\subsection{Standard Fields}

We defined our primary statistical samples as the union of the NDWFS field geometry with
a set of 15 sub-fields defined by the Hectospec field of view.  Sources had to
lie within $0.49$~deg of one of the 15 field centers illustrated in Fig.~\ref{fig:fields} and 
listed in Table~\ref{tab:fields}.  The field centers were simply defined by the final field centers
used for the primary observational runs in 2004.  Sources inside the standard sub-fields
are assigned the field ID of the closest field center, while those outside are assigned
a field ID of $-1$ (see Table~\ref{tab:passes}). 
Several of the circular fields extend beyond the NDWFS area, so the actual survey area
must be clipped to exclude $\hbox{RA}<228.96$~deg, $\hbox{Dec} > 33.46$~deg, and $\hbox{Dec} < 35.84$~deg
if $\hbox{RA}>216.14$~deg.    
The total area within this region is $2.40\times 10^{-3}$~sr
(7.88~deg$^2$).  We also do not target sources within radius $R_{bstar}/2$ of a 
bright $R_{USNO}<17$~mag USNO star, where $R_{bstar} = 20\farcs0 + 5\farcs0(15-R_{USNO})$.
For $R_{bstar}/2 < R < R_{bstar}$, there were additional surface brightness criteria for
observing targets (see below). Excluding the bright star exclusion areas with $R<R_{bstar}/2$,
the survey area drops to $2.36 \times 10^{-3}$~sr (7.74~deg$^2$).

The next three sub-sections describe the evolving selection criteria.  Most readers 
should proceed to \S\ref{sec:finalsample} which defines the final sample selection
criteria.  The criteria for the earlier observations in \S\ref{sec:firstsample}
and \S\ref{sec:middlesample} are supplied for completeness and because all objects
targeted in these earlier seasons continued to be targeted in the later seasons
independent of any revisions to the selection criteria.  Tables \ref{tab:samples1}
and \ref{tab:samples2} summarize the final samples and their completeness.  

\subsection{2004 Sample Definitions}
\label{sec:firstsample}

The 2004 samples were based on preliminary NDWFS and IRAC Shallow Survey catalogs and
photometric calibrations.  The observations also immediately followed the initial 
engineering runs for Hectospec.  Since there was no experience
with the performance of Hectospec in the red, we decided to set our optical selection
criterion using a catalog\footnote{NDWFS\_R2150\_rot\_apr08\_newcat}
to an R-band magnitude limit of $\rmag<21.5$ so that K-corrections would minimize the number of
galaxies with $z>0.5$.   
The NDWFS optical photometry was flagged as good if the Kron-like, mag$\_$auto magnitude
was defined (magnitude between 0 and 80), the SExtractor flags were $\hbox{FLAGS}<8$, 
catalog duplication flag $\hbox{FLAG\_DUPLICATE}=0$ and it was detected in
more than one sub-images available for the band.  A galaxy ($\hbox{galaxy}=1$) 
was required to have good data in R and either I or B$_W$, SExtractor stellarity indices
$\leq0.8$ in all bands, $\rmag\leq 20$~mag and $\rmag_{ap1}\leq 23.5$~mag.  
Star/galaxy separation in the NDWFS catalogs based on the SExtractor stellarity 
indices is effective to significantly fainter fluxes than our spectroscopic flux
limits.  We explicitly included all galaxies found in the 2MASS survey and excluded all galaxies 
within radius $20\farcs0-R_{USNO}$ of a USNO star with $R_{USNO} \leq 17$~mag.  A
quasar target needed to have $17 < \rmag \leq 21$, good R-band data, and not
have $\hbox{galaxy} = 1$.
There were 15 target groups defined for 2004 defined by the 
binary target code $\hbox{code04}$. The complete, main R-band galaxy sample
is the combination of $\hbox{code04}=2048$ (bright R-band galaxies) and
$\hbox{code04}=512$ (20\% sparse sampling of fainter R-band galaxies).   
After the first series of observations, we
were beginning to exhaust the AGN targets, so we added the fainter X-ray and
MIPS target categories as well as a set of experimental brown dwarf candidates ($\hbox{code04}=4096$,
$8192$ and $16384$) at lower priority.  Because these observations were significantly
dependent on preliminary photometry, we include the targeting information but not 
the underlying photometry in this paper.  The 15 samples are:

\begin{itemize}
\item{SDSS Flux Calibration Stars ($\hbox{code04}=1$):} These are candidate F-stars selected on the basis of 
   SDSS photometry that are used to flux calibrate the spectra.  We tried to include
   5 of these flux calibration stars in each observation.
\item{IRAC 8.0$\mu$m galaxy sample ($\hbox{code04}=2$):}
   All galaxies ($\hbox{galaxy}=1$) with $[8.0]_{[3.6]}(6\farcs0)\leq 13.2$~mag and $\rmag\leq20$~mag. 
   As a reminder, all four IRAC samples were based on preliminary versions of the 
   IRAC Shallow Survey catalogs.
\item{IRAC 5.8$\mu$m galaxy sample ($\hbox{code04}=4$):}
   All galaxies with $[5.8]_{[3.6]}(6\farcs0)\leq 14.7$~mag and $\rmag\leq20$~mag. 
\item{IRAC 4.5$\mu$m galaxy sample ($\hbox{code04}=8$):}
   All galaxies with $[4.5]_{[3.6]}(6\farcs0)\leq 15.2$~mag and $\rmag\leq20$~mag. 
\item{IRAC 3.6$\mu$m galaxy sample ($\hbox{code04}=16$):}
   All galaxies with $[3.6]_{[3.6]}(6\farcs0)\leq 15.2$~mag and $\rmag\leq20$~mag.  
\item{MIPS 24$\mu$m Sources ($\hbox{code04}=32$):} These targets were galaxies or point sources with 
   $F_{24}\geq 1$~mJy with optical flux limits of $R\leq 20$ for galaxies and $R \leq 21.5$ for
   stellar targets and $R_{ap1}<23.5$ for both. Stellar targets also had to either lack
   2MASS detections or have $\hbox{J} > 12 - 2.5\log(F_{24}/\hbox{mJy})$~mag in
   order to eliminate normal stars.  We will illustrate this criterion for the later
   seasons where we used an I-band variant of this criterion. 
\item{Blue Galaxy Sample ($\hbox{code04}=64$):} This sample consisted of all galaxies with 
    $\bmag<20.5$~mag, $\hbox{bgood}=1$ and $\rmag\leq20$~mag.
\item{Compact FIRST Sources ($\hbox{code04}=128$):} This sample consists of FIRST radio sources
    with deconvolved axes smaller than 1\farcs0 whose positions were within $3\farcs0$
    of an $R \leq 21.5$ and $R_{ap1}\leq 23.5$ optical source.  Here, and in the X-ray
    samples, this latter criterion was to ensure that the flux in a Hectospec fiber was
    large enough to plausibly measure the redshift.
\item{Bright X-ray Quasar Candidates ($\hbox{code04}=256$):} This sample consists of sources from
    the XBo\"otes catalog with 4 or more X-ray counts that were matched to the $\rmag\leq 21.5$
    optical catalog and also have $R_{ap1}<23.5$, independent of whether they were extended or
    stellar sources.
\item{Main Faint R-band galaxy sample ($\hbox{code04}=512$):} This sample consists of a randomly selected
    20\% ($\hbox{rcode}\leq 3$) of galaxies with $19.2<\rmag\leq20$~mag.  The complete main
    R-band galaxy sample consists of this sub-sample plus the bright R-band galaxy sample ($\hbox{code04}=2048$).
\item{Faint R-band galaxies ($\hbox{code04}=1024$):} This sample consists of all galaxies
    with $19.2<\rmag\leq 20$~mag.  The first 20\% of these galaxies ($\hbox{rcode}\leq 3$)
    are part of the Main Faint R-band Galaxy Sample ($\hbox{code04}=512$) as well, and are 
    observed at high priority.  The remaining galaxies were observed with priorities that
    favored lower rcodes over higher rcodes.
\item{Main Bright R-band galaxy sample ($\hbox{code04}=2048$):} This sample consists of all galaxies
    with $\rmag\leq19.2$~mag.
\item{Fainter X-ray sources ($\hbox{code04}=4096$)} These are fainter (2 or 3 count) sources from the
    XBo\"otes survey. Otherwise the criteria were the same as for the main X-ray sample.
    While targeting 2 count X-ray sources sounds odd, the backgrounds of the XBo\"otes survey
    are so low that almost all such sources associated with optical sources brighter than the 
    spectroscopic flux limits will be real. 
\item{Fainter MIPS point sources ($\hbox{code04}=8192$):}  These were point sources with
    $0.5 \leq F_{24}\leq 1.0$~mJy that otherwise satisfied the point source criteria
    for the main MIPS sample.  Galaxies were not included here.
\item{IRAC brown dwarf candidates ($\hbox{code04}=16384$):} These targets were supplied by M. Ashby as
    an experiment, and were all found to be star forming galaxies.  Since they are not part of the primary
    AGES samples, we include them without further discussion because they were a low priority targeting criterion. 
\end{itemize}

\begin{figure}[p]
\centerline{\includegraphics[width=5.0in]{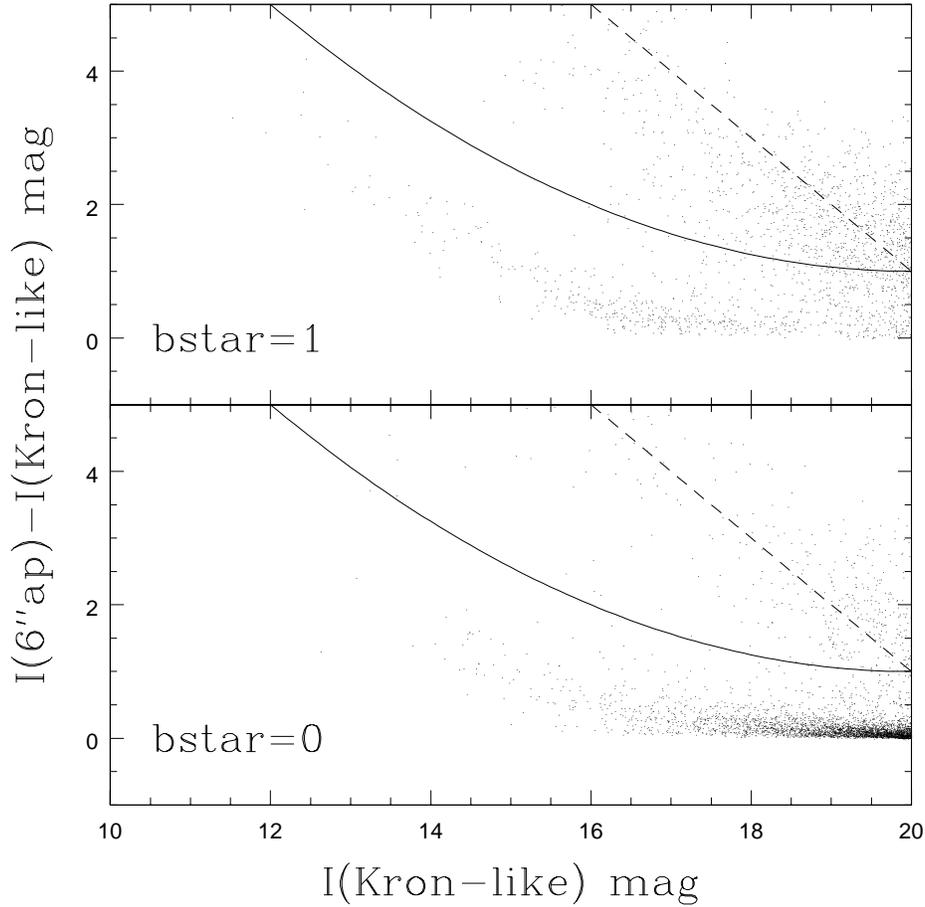}}
\caption{ NDWFS bright star photometry problems.  The top panel shows the difference between the $\hbox{I}(6\farcs0)$
   aperture magnitude and the Kron-like (mag$\_$auto) $\hbox{I}$ magnitudes for 5000 randomly selected galaxies near
   bright USNO stars ($\hbox{bstar}=1$). The bottom panel shows the same quantities for 5000 galaxies
   which are not ($\hbox{bstar}=0$).  The Kron-like I band magnitudes tend to be overestimated when 
   the source is close to a bright star.  This is not true of the R band magnitudes used in the first
   season.  All galaxies are required to have $I_{ap6}<21$~mag, which is 
   indicated by the dashed line, while those close to bright stars are eliminated if $r<R_{bstar}/2$ 
   or if $R_{bstar}/2< r < R_{bstar}$ and they lie above the solid line, $I_{ap6} > I + 1 + 4\left[(I-20)/8\right]^2$~mag. 
   }
\label{fig:bstar}
\end{figure}

\subsection{2005 Sample Definitions}
\label{sec:middlesample}

The 2005 sample definitions were very different from those in 2004 because
the primary optical band was changed from the R-band to the I-band.  It was clear at
this point that Hectospec would work well at our desired flux levels as the 4000\AA\
break moved beyond the R band, and we wanted the evolutionary leverage from pushing the
typical redshift upwards that would be gained from using an I-band flux limit.  
We started with all objects having $\imag \leq 21.5$~mag in the NDWFS 
DR3 catalogs and then matched them to all the other bands.  
The NDWFS optical photometry was flagged as good if the Kron-like, mag$\_$auto magnitude
was defined (magnitude between 0 and 80), the SExtractor flags were $\hbox{FLAGS}<8$, the
catalog duplication flag was $\hbox{FLAG\_DUPLICATE}=0$ and photometric
data was available ($\hbox{FLAG\_PHOT}=1$).  
An object was 
defined as a point source ($\hbox{pntsrc}=1$) if it had a SExtractor stellarity index
$\geq 0.8$ in any of the B$_W$, R, or I-bands.  An object was a good target ($\hbox{good}=1$)
if $\hbox{igood}=1$ and either $\hbox{bgood}$ or $\hbox{rgood}=1$.
Galaxy targets ($\hbox{galaxy}=1$) were good ($\hbox{good}=1$), extended ($\hbox{pntsrc}=0$) targets
with $\hbox{I}\leq 20$~mag, $\imag_{ap1}\leq 24$ and $\imag_{ap6} \leq 21$~mag.  
Quasar targets ($\hbox{qso}=1$) were good ($\hbox{good}=1$), point sources ($\hbox{pntsrc}=1$) with
$\hbox{I}\leq 21.5$~mag and $\imag_{ap1}\leq 24$.  We only attempted to obtain redshifts
for galaxies and quasars with $\imag>15$ and $16$~mag respectively.  The redshifts of
brighter sources were filled in using SDSS (e.g. DR7, \citealt{Abazajian2009}).  We also used the final rather
than the preliminary versions of the IRAC Shallow Survey catalogs (\citealt{Eisenhardt2004}) and switched
to using the Kron-like magnitudes ($\hbox{[3.6]}_{[3.6]} \cdots \hbox{[8.0]}_{[8.0]}$) rather than the 
6\farcs0 aperture magnitudes ($\hbox{[3.6]}_{[3.6]}(6\farcs0) \cdots \hbox{[8.0]}_{[8.0]}(6\farcs0)$).
Only a small portion of the NDWFS field had been observed by GALEX at this point, and
the GALEX UV-selected galaxy samples based on these preliminary catalogs
probably should not be directly used.

The Kron-like I-band SExtractor magnitudes clearly had significantly more problems near bright
stars than the R-band magnitudes used in 2004, as shown in Fig.~\ref{fig:bstar}.  We flagged galaxies as being potentially 
affected by bright stars ($\hbox{bstar}=1$) if they lay within the magnitude dependent
radius $R_{bstar} =20\farcs0+5\farcs0(15-R_{USNO})$ of a star with an R-band USNO magnitude 
$R_{USNO}\leq17$~mag.  Galaxies with $\hbox{bstar}=1$ were rejected ($\hbox{galaxy}=0$)
if they had a surface brightness $\imag_{ap6}>\imag+4[(I-20)/8]^2$~mag or they were
within $R_{bstar}/2$ of a bright star.  Eqn.~1 in \S6 gives a procedure from \cite{Cool2011}
for controlling this problem.
 
There were 20 sub-samples in the 2005 survey definition.  Galaxy samples were now
defined at the GALEX FUV and NUV, I-band, J-band and K-band as well as the B$_W$,
R, IRAC and MIPS bands.  Quasar samples were now defined in the IRAC and optical
bands in addition to the X-ray, MIPS and radio targeting, and we used the WSRT 
radio sources rather than FIRST.  Since the full $\hbox{code05}$ values were
becoming unwieldy, we also assigned sub-codes for galaxy ($\hbox{gcode05}$)
and quasar ($\hbox{qcode05}$)  samples where $\hbox{code05}=\hbox{qcode05}+128\times\hbox{gcode05}$.

\begin{figure}[p]
\centerline{\includegraphics[width=5.0in]{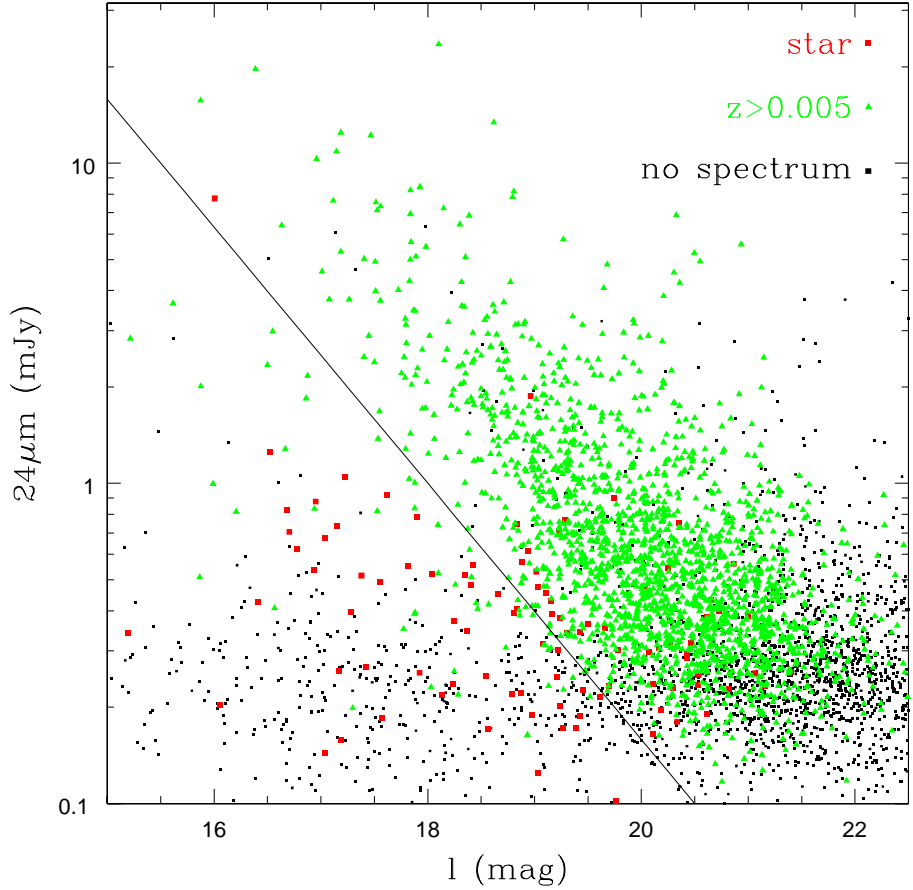}}
\caption{ MIPS quasar selection.  MIPS quasar targets are point sources with $\imag(3\farcs0) > 18-2.5\log(F_{24}/\hbox{mJy})$. 
   The green filled triangles show extragalactic sources, the filled red squares show stars, and the black squares are
   sources without spectroscopic confirmations.  The black line indicates the color selection boundary.  
   Targets appear below the line because of other targeting criteria.
   }
\label{fig:mipsqso}
\end{figure}

\begin{itemize}
\item{SDSS Calibration Stars: ($\hbox{code05}=1$):} These are SDSS stars with
the colors of F stars that are used to flux calibrate the spectra.

\item{Brown Dwarf Candidates: ($\hbox{code05}=2$, $\hbox{qcode05}=1$):} These
are the same brown dwarf candidates as in 2004, and we do not discuss them
further.

\item{Optical Quasar Candidates: ($\hbox{code05}=4$, $\hbox{qcode05}=2$):}  These
are B$_W$/R/I/K-band color-selected quasar candidates from an experiment by
K. Brand and R. Green.  The first class of targets consists of point sources with 
$\imag-\kmag \geq 0.5 + (4.0/5.8)(\bmag-\rmag) $ or $\imag-\kmag>3$.  The
second class of objects consist of B$_W$ non-detections which satisfy one of
$\rmag-\imag < 1.0$, $\imag-\kmag > 1.1 + (\rmag-\imag)$ or $\imag-\kmag > 3.0$.
This was a small sample designed to test the color selection method and we
do not discuss it further.

\item{WSRT Radio Sources: ($\hbox{code05}=8$, $\hbox{qcode05}=4$):}  All sources
($\hbox{qso}=1$ or $\hbox{galaxy}=1$) within 3\farcs0 of a $5\sigma$ detection in
the WSRT 1.4 GHz survey of the field (\citealt{deVries2002}).  We made no attempt
to deal with the problem of radio lobes other than to select unresolved sources
in the \cite{deVries2002} catalogs.

\item{X-ray Quasar Candidates: ($\hbox{code05}=16$, $\hbox{qcode05}=8$):}  All sources
($\hbox{qso}=1$ or $\hbox{galaxy}=1$) with 2 or more X-ray counts and a greater than
25\% Bayesian probability of being identified with the optical source using the
matching approach outlined in \cite{Brand2006}.  Remember that the optical flux
limits are different for the extended and point-like targets.

\item{MIPS Quasar Candidates: ($\hbox{code05}=32$, $\hbox{qcode05}=16$):}  All point
sources with $F_{24}\geq 0.3$~mJy and $\imag(3\farcs0) > 18-2.5\log(F_{24}/\hbox{mJy})$.
We changed to using an I-band/24$\mu$m criterion to eliminate stars
rather than a J-band/24$\mu$m criterion.  Also note that the 
MIPS flux limit is below the 80\% completeness limit of the 
24$\mu$m catalogs. Fig.~\ref{fig:mipsqso} illustrates this selection method.

\item{IRAC Quasar Candidates: ($\hbox{code05}=64$, $\hbox{qcode05}=32$):}  This sample
includes both galaxies and point sources, with the standard optical flux limits
of $\imag\leq 20$ for the extended sources and $\imag \leq 21.5$ for the point sources.
The selection criteria are based on \cite{Stern2005}, but have been modified to be
more liberal for point sources and slightly more conservative for extended sources.

\begin{itemize}
   \item Point sources:  Point sources brighter than $[3.6]_{[3.6]}\leq 18$~mag only needed to
      satisfy the criterion that $[3.6]_{[3.6]}(3\farcs0)-[4.5]_{[4.5]}(3\farcs0)\geq 0.4$~mag because
      we need only differentiate quasars from stars with (Vega) mid-IR colors of zero.
      For fainter point sources, where the color errors become significant, we added
      the requirement that the source had a measured $[5.8]_{[5.8]}(3\farcs0)$ or $[8.0]_{[8.0]}(3\farcs0)$
      magnitude.

   \item Extended Sources:  Extended sources had to be detected in all four IRAC bands
      and satisfy the IRAC color cuts (these are all 3\farcs0 aperture magnitudes)
      \begin{itemize}
        \item
        $[5.8]_{[5.8]}-[8.0]_{[8.0]} > 0.6$,
        \item
        $[3.6]_{[3.6]}-[4.5]_{[4.5]}>0.2 \left([5.8]_{[5.8]}-[8.0]_{[8.0]}\right)+0.4$ and
        \item
        $[3.6]_{[3.6]}-[4.5]_{[4.5]}>2.5\left([5.8]_{[5.8]}-[8.0]_{[8.0]}\right)-3.5$.
      \end{itemize}
\end{itemize}

\item{MIPS 24$\mu$m Galaxy Sample: ($\hbox{code05}=128$, $\hbox{gcode05}=1$):}  This sample
consists of galaxies ($\hbox{galaxy}=1$) with $F_{24}\geq 0.3$~mJy.  We attempted to obtain redshifts of all 
galaxies with $F_{24}\geq 0.5$~mJy and a randomly selected 30\% ($\hbox{rcode}\leq 5$) of the 
galaxies with $0.3 \leq F_{24} <0.5$~mJy.  Note that these 24$\mu$m flux limits are fainter 
than the 80\% completeness limit of the 24$\mu$m catalogs.  

\item{IRAC Channel 4 (8.0$\mu$m) Galaxy Sample: ($\hbox{code05}=256$, $\hbox{gcode05}=2$):}  This sample
consists of galaxies ($\hbox{galaxy}=1$) with $[8.0]_{[8.0]} \leq 13.8$~mag.  We attempted to obtain redshifts of all 
galaxies with $[8.0]_{[8.0]}\leq 13.2$~mag and a randomly selected 30\% ($\hbox{rcode}\leq 5$) of the 
galaxies with $13.2 < [8.0]_{[8.0]} \leq 13.8$~mag.  

\item{IRAC Channel 3 (5.8$\mu$m) Galaxy Sample: ($\hbox{code05}=512$, $\hbox{gcode05}=4$):}  This sample
consists of galaxies ($\hbox{galaxy}=1$) with $[5.8]_{[5.8]} \leq 15.2$~mag.  We attempted to obtain redshifts of all 
galaxies with $[5.8]_{[5.8]}\leq 14.7$~mag and a randomly selected 30\% ($\hbox{rcode}\leq 5$) of the 
galaxies with $14.7 < [5.8]_{[5.8]} \leq 15.2$~mag.  

\item{IRAC Channel 2 (4.5$\mu$m) Galaxy Sample: ($\hbox{code05}=1024$, $\hbox{gcode05}=8$):}  This sample
consists of galaxies ($\hbox{galaxy}=1$) with $[4.5]_{[4.5]} \leq 15.7$~mag.  We attempted to obtain redshifts of all 
galaxies with $[4.5]_{[4.5]}\leq 15.2$~mag and a randomly selected 30\% ($\hbox{rcode}\leq 5$) of the 
galaxies with $15.2 < [4.5]_{[4.5]} \leq 15.7$~mag.  

\item{IRAC Channel 1 (3.6$\mu$m) Galaxy Sample: ($\hbox{code05}=2048$, $\hbox{gcode05}=16$):}  This sample
consists of galaxies ($\hbox{galaxy}=1$) with $[3.6]_{[3.6]} \leq 15.7$~mag.  We attempted to obtain redshifts of all 
galaxies with $[3.6]_{[3.6]}\leq 15.2$~mag and a randomly selected 30\% ($\hbox{rcode}\leq 5$) of the 
galaxies with $15.2 < [3.6]_{[3.6]} \leq 15.7$~mag.  

\item{GALEX FUV-band Galaxy Sample: ($\hbox{code05}=4096$, $\hbox{gcode05}=32$):}  This sample
consists of galaxies ($\hbox{galaxy}=1$) with $\hbox{FUV} \leq 22.5$~mag.  
We attempted to obtain redshifts of all 
galaxies with $\hbox{FUV}\leq 22.0$~mag and a randomly selected 30\% ($\hbox{rcode}\leq 5$) of the 
galaxies with $22.0 < FUV \leq 22.5$~mag.  The GALEX data available at the time covered only a
small fraction of the standard fields.  

\item{GALEX NUV-band Galaxy Sample: ($\hbox{code05}=8192$, $\hbox{gcode05}=64$):}  This sample
consists of galaxies ($\hbox{galaxy}=1$) with $\hbox{NUV} \leq 22.0$~mag.  
We attempted to obtain redshifts of all 
galaxies with $\hbox{NUV}\leq 21.0$~mag and a randomly selected 30\% ($\hbox{rcode}\leq 5$) of the 
galaxies with $21.0 < NUV \leq 22.0$~mag.  

\item{K-band Galaxy Sample: ($\hbox{code05}=16384$, $\hbox{gcode05}=128$):}  This sample
consists of galaxies ($\hbox{galaxy}=1$) with either NDWFS K/K$_s$ or FLAMEX $\kmag_s \leq 16.5$~mag.  
We attempted to obtain redshifts of all 
galaxies with $\kmag\leq 16.0$~mag and a randomly selected 20\% ($\hbox{rcode}\leq 3$) of the 
galaxies with $16.0 < \kmag \leq 16.5$~mag.

\item{J-band Galaxy Sample: ($\hbox{code05}=32768$, $\hbox{gcode05}=256$):}  This sample
consists of galaxies ($\hbox{galaxy}=1$) with FLAMEX $\jmag \leq 18.5$~mag.  We attempted to obtain redshifts of all 
galaxies with $\jmag\leq 17.5$~mag and a randomly selected 20\% ($\hbox{rcode}\leq 3$) of the 
galaxies with $17.5 < \bmag \leq 18.5$~mag.

\item{B$_W$-band Galaxy Sample: ($\hbox{code05}=65536$, $\hbox{gcode05}=512$):}  This sample
consists of galaxies ($\hbox{galaxy}=1$) with $\bmag \leq 21.3$.  We attempted to obtain redshifts of all 
galaxies with $\bmag\leq 20.5$~mag and a randomly selected 20\% ($\hbox{rcode}\leq 3$) of the 
galaxies with $20.5 < \bmag \leq 21.3$~mag.  The bright ($\bmag<20.5$) part of this sample should 
be very similar to the 2004 B$_W$-band galaxy sample ($\hbox{code04}=64$).

\item{R-band Galaxy Sample: ($\hbox{code05}=131072$, $\hbox{gcode05}=1024$):}  This sample
consists of galaxies ($\hbox{galaxy}=1$) with $\rmag \leq 20$.  We attempted to obtain redshifts of all 
galaxies with $\rmag\leq 19.2$~mag and a randomly selected 20\% ($\hbox{rcode}\leq 3$) of the 
galaxies with $19.2 < \rmag \leq 20$~mag.  This sample should be very similar to the 2004
R-band galaxy sample ($\hbox{code04}=2048$ plus $\hbox{code04}=512$). 

\item{Other I-band Galaxies: ($\hbox{code05}=262144$, $\hbox{gcode05}=2048$):}  This sample
consists of all galaxies ($\hbox{galaxy}=1$) with $\imag \leq 20$~mag that were not included in
the Main I-band Galaxy sample.  These sources were observed at lower priority than the main samples.
Galaxies with lower $\hbox{rcode}$ values are preferentially observed to make it easier for any
later survey to produce larger randomly selected sub-samples. 

\item{Main I-band Galaxy Sample: ($\hbox{code05}=524288$, $\hbox{gcode05}=4096$):}  This sample
consists of galaxies ($\hbox{galaxy}=1$) with $\imag \leq 20$.  We attempted to obtain redshifts of all 
galaxies with $15\leq\imag\leq 18.5$~mag and a randomly selected 20\% ($\hbox{rcode}\leq 3$) of the 
galaxies with $18.5 < \imag \leq 20$~mag.

\end{itemize}  

\subsection{2006 and 2007 Sample Definitions}
\label{sec:finalsample}

The 2006 sample definitions are very similar to those of 2005 except for 
changes in the AGN sample definitions to make use of the zBo\"otes data and to
better characterize selection effects.  The 2006 sample definitions were used
again in 2007.  The basic sample was selected from the I band catalog and
then matched to all the other bands.  The NDWFS optical photometry was flagged 
as good (bgood, rgood or $\hbox{igood}=1$)
if the Kron-like magnitude
was defined (magnitude between 0 and 80), the SExtractor flags were $\hbox{FLAGS}<8$ and the
catalog duplication flag was $\hbox{FLAG\_DUPLICATE}=0$.  The criterion that photometric
data was available ($\hbox{FLAG\_PHOT}=1$) was dropped.  For the z'-band, objects were
flagged as good ($\hbox{zgood}=1$) if the source was not split, came from a region with 
more than 4 observations and was
not flagged in the zBo\"otes catalog as being near a bright star.  An object was 
defined as a point source, $\hbox{pntsrc}=1$, if it had a SExtractor stellarity index
$\geq 0.8$ in any of the B$_W$, R, I or z'-bands with good data ($\hbox{bgood}=1$ etc).   

A galaxy target ($\hbox{galaxy}=1$) was required to have $\hbox{pntsrc}=0$, $\hbox{igood}=1$ and one 
of rgood, bgood or $\hbox{zgood}=1$.  It then had to satisfy the (Kron-like) I-band magnitude criteria 
$\imag\leq20$~mag, 1\farcs0 aperture magnitude $\imag_{ap1}\leq 24.0$ and 6\farcs0 aperture magnitude 
$\imag_{ap6}\leq 21.0$.  A quasar target ($\hbox{qso}=1$) had to have either $\hbox{igood}=1$ or $\hbox{zgood}=1$, 
which is more liberal than in 2005 because
requirements on rgood or bgood could be problematic for very high redshift quasars.  It then had to satisfy either
that $\imag \leq 22.5$ and a 1\farcs0 aperture I-band magnitude $\imag_{ap1}\leq 24$~mag or that $\zmag\leq22.5$ and a  
1\farcs0 z-band magnitude $\zmag_{ap1} \leq 24.0$~mag.  We also included a separate category AGN/galaxy 
($\hbox{agngalaxy}=1$) 
which was an extended source that did not have to meet the criterion on the 6\farcs0 aperture magnitude and 
included sources down to the faint limit used for the point sources $\imag\leq22.5$~mag rather than the limit 
used for normal galaxies of $\imag\leq20$~mag.   The limit on the aperture
magnitude is designed to filter out problems created by bright stars. 

\begin{itemize}

\item{SDSS Calibration Stars: ($\hbox{code06}=1$):} These are SDSS stars with
the colors of F stars that are used to flux calibrate the spectra.

\item{Brown Dwarf Candidates: ($\hbox{code06}=2$, $\hbox{qcode06}=1$):} These
are brown dwarf candidates (M. Ashby, private communication).
This sample
is unchanged from 2005 other than through the modified definitions of galaxies and point sources.

\item{Optical Quasar Candidates: ($\hbox{code06}=4$, $\hbox{qcode06}=2$):}  These
are B$_W$/R/I/K-band color-selected quasar candidates (Brand \& Green, private
communication).  This sample is unchanged from 2005 other than through the modified 
definitions of galaxies and point sources.
  
\item{WSRT Radio Sources: ($\hbox{code06}=8$, $\hbox{qcode06}=4$):}  All sources
($\hbox{qso}=1$, $\hbox{galaxy}=1$, or $\hbox{agngalaxy}=1$) within 3\farcs0 of a $5\sigma$ detection in
the WSRT 1.4 GHz survey of the field (de Vries et al. 2002).
This differs from 2005 by
including the faint, extended sources with $\hbox{agngalaxy}=1$.

\item{X-ray Quasar Candidates: ($\hbox{code06}=16$, $\hbox{qcode06}=8$):}  All sources
($\hbox{qso}=1$, $\hbox{galaxy}=1$, or $\hbox{agngalaxy}=1$) with 2 or more X-ray counts and a greater than
25\% Bayesian probability of being identified with the optical source using the
matching approach outlined in \cite{Brand2006}.  This differs from 2005 by
including the faint, extended sources with $\hbox{agngalaxy}=1$.

\item{MIPS Quasar Candidates: ($\hbox{code06}=32$, $\hbox{qcode06}=16$):}  
This sample is unchanged from 2005 other than through the modified definitions
of galaxies and point sources.

\item{IRAC Quasar Candidates: ($\hbox{code06}=64$, $\hbox{qcode06}=32$):}  This sample is the most
heavily modified from 2005.  The changes were implemented to better understand selection effects
due to color and morphology.  For point sources, it was clear from detailed analyses that the old
color criterion led to reduced completeness whenever a bright emission line was in the 3.6$\mu$m
band, in particular at $z\sim 4.5$ with the H$\alpha$ line (see \citealt{Assef2010}).  It was also clear that the differing
magnitude limits for point and extended sources were a significant problem at low redshifts.
The new selection criterion were sufficiently complex that we introduced
a separate code ($\hbox{iracq06}$) to label the various criteria.  We also switched to using
colors measured with positions set by the 3.6$\mu$m band ($[x]_{[3.6]}$ magnitudes rather than $[x]_x$
magnitudes).

\begin{itemize}
   \item{$\hbox{iracq06}=1$:} Point sources brighter than $[3.6]_{[3.6]}\leq 18$~mag with
      $[3.6]_{[3.6]}(3\farcs0)-[4.5]_{[3.6]}(3\farcs0)\geq 0.4$~mag.  This differs from 2005 only
      in using the same band to set the position of the aperture by using $[4.5]_{[3.6]}$
      instead of $[4.5]_{[4.5]}$.

   \item{$\hbox{iracq06}=2$:} Point sources in the magnitude range $18.0<[3.6]_{[3.6]} \leq 18.5$~mag with
      $[3.6]_{[3.6]}(3\farcs0)-[4.5]_{[3.6]}(3\farcs0)\geq 0.4$~mag and either
      $\imag(3\farcs0)-[3.6]_{[3.6]}(3\farcs0)\geq 3$ or $\zmag(3\farcs0)-[3.6]_{[3.6]}(3\farcs0)\geq 3$.
      The latter two criteria were included to minimize stellar contamination without
      requiring a detection in the less sensitive $5.8$ or $8.0\mu$m IRAC bands.

   \item{$\hbox{iracq06}=3$:} The point source criteria for $\hbox{irac06}=1$ or $2$
      can lose objects when strong emission lines pass through the 3.6$\mu$m band,
      particularly at $z\sim 4.5$ (H$\alpha$).  To minimize stellar contamination, 
      we added objects with $[3.6]_{[3.6]} \leq 18.0$ and $[3.6]_{[3.6]}(3\farcs0)-[4.5]_{[3.6]}(3\farcs0)\geq 0.3$ 
      mag subject to the added criterion on the optical to mid-IR color that either
      $\imag(3\farcs0)-[3.6]_{[3.6]}(3\farcs0)\geq 3$ or $\zmag(3\farcs0)-[3.6]_{[3.6]}(3\farcs0)\geq 3$.  
      We included 25\% ($\hbox{rcode}\leq 4$) of these sources at 
      high priority, and the remainder at low priority.

   \item{$\hbox{irac06}=4$:} Extended sources flagged as standard galaxies ($\hbox{galaxy}=1$,
      so $\imag \leq 20$) had to be detected in all four IRAC bands
      and satisfy the IRAC color cuts
      \begin{itemize}
      \item
      $[5.8]_{[3.6]}(3\farcs0)-[8.0]_{[3.6]}(3\farcs0) > 0.6$,
      \item
      $[3.6]_{[3.6]}(3\farcs0)-[4.5]_{[3.6]}(3\farcs0)>0.2\left([5.8]_{[3.6]}(3\farcs0)-[8.0]_{[8.0]}(1\farcs0)\right)+0.4$ and
      \item
      $[3.6]_{[3.6]}(3\farcs0)-[4.5]_{[3.6]}(3\farcs0)>2.5\left([5.8]_{[3.6]}(3\farcs0)-[8.0]_{[8.0]}(1\farcs0)\right)-3.5$.
      \end{itemize}
      This differs from 2005 only in using $[x]_{[3.6]}$ rather than $[x]_x$ to define the
      colors.

   \item{$\hbox{irac06}=5$:} Extended sources that were not flagged as standard galaxies
      ($\hbox{galaxy}=1$ and thus extended sources with $\imag \leq 20$), but were flagged as faint galaxies
      ($\hbox{agngalaxy}=1$, so extended sources $20 \leq \imag \leq 22.5$), were detected in all four IRAC bands
      and satisfied the IRAC color cuts
      \begin{itemize}
      \item
      $[5.8]_{[3.6]}(3\farcs0)-[8.0]_{[3.6]}(3\farcs0) > 0.6$,
      \item
      $[3.6]_{[3.6]}(3\farcs0)-[4.5]_{[3.6]}(3\farcs0)>0.2\left([5.8]_{[3.6]}(3\farcs0)-[8.0]_{[3.6]}(1\farcs0)\right)+0.4$ and
      \item
      $[3.6]_{[3.6]}(3\farcs0)-[4.5]_{[3.6]}(3\farcs0)>2.5\left([5.8]_{[3.6]}(3\farcs0)-[8.0]_{[3.6]}(1\farcs0)\right)-3.5$.
      \end{itemize}

\end{itemize}

\item{MIPS 24$\mu$m Galaxy Sample: ($\hbox{code06}=128$, $\hbox{gcode06}=1$):}  This sample
is unchanged from 2005 other than through the modified definitions of galaxies and point sources.

\item{IRAC Channel 4 (8.0$\mu$m) Galaxy Sample: ($\hbox{code06}=256$, $\hbox{gcode06}=2$):}  This sample
is unchanged from 2005 other than through the modified definitions of galaxies and point sources.

\item{IRAC Channel 3 (5.8$\mu$m) Galaxy Sample: ($\hbox{code06}=512$, $\hbox{gcode06}=4$):}  This sample
is unchanged from 2005 other than through the modified definitions of galaxies and point sources.

\item{IRAC Channel 2 (4.5$\mu$m) Galaxy Sample: ($\hbox{code06}=1024$, $\hbox{gcode06}=8$):}  This sample
is unchanged from 2005 other than through the modified definitions of galaxies and point sources.

\item{IRAC Channel 1 (3.6$\mu$m) Galaxy Sample: ($\hbox{code06}=2048$, $\hbox{gcode06}=16$):}  This sample
is unchanged from 2005 other than through the modified definitions of galaxies and point sources.

\item{GALEX FUV-band Galaxy Sample: ($\hbox{code06}=4096$, $\hbox{gcode06}=32$):}  
This sample
was rebuilt from the public GALEX catalogs available for the field in 2007, within $0.45$~deg of the
GALEX field center and with at least $2000$~sec of NUV integration time.  The GALEX data still
covered only a modest fraction of the standard fields. 

\item{GALEX NUV-band Galaxy Sample: ($\hbox{code06}=8192$, $\hbox{gcode06}=64$):}  
This sample
was rebuilt from the public GALEX catalogs available for the field in 2007, within $0.45$~deg of the
GALEX field center and with at least $2000$~sec of NUV integration time.

\item{K-band Galaxy Sample: ($\hbox{code06}=16384$, $\hbox{gcode06}=128$):}  This sample
is unchanged from 2005 other than through the modified definitions of galaxies and point sources.

\item{J-band Galaxy Sample: ($\hbox{code06}=32768$, $\hbox{gcode06}=256$):}  This sample
is unchanged from 2005 other than through the modified definitions of galaxies and point sources.

\item{B$_W$-band Galaxy Sample: ($\hbox{code06}=65536$, $\hbox{gcode06}=512$):}  This sample
is unchanged from 2005 other than through the modified definitions of galaxies and point sources.

\item{R-band Galaxy Sample: ($\hbox{code06}=131072$, $\hbox{gcode06}=1024$):}  This sample
is unchanged from 2005 other than through the modified definitions of galaxies and point sources.

\item{Other I-band Galaxies: ($\hbox{code06}=262144$, $\hbox{gcode06}=2048$):}  This sample
is unchanged from 2005 other than through the modified definitions of galaxies and point sources.

\item{Main I-band Galaxy Sample: ($\hbox{code06}=524288$, $\hbox{gcode06}=4096$):}  This sample
is unchanged from 2005 other than through the modified definitions of galaxies and point sources.

\end{itemize}

\section{Observations}
\label{sec:observation}

The observations were made with Hectospec (\citealt{Fabricant1998}, 
\citealt{Roll1998}, \citealt{Fabricant2005}), 
a 300 fiber, 1~degree field of view, robotic spectrograph
for the 6.5m MMT telescope at Mt. Hopkins.  The wavelength range is 3700\AA\ to 9200\AA\ with a pixel
scale of 1.2\AA\ and a spectral resolution of 6\AA\ (i.e. roughly $R\sim 1000$).  
The throughput at the wavelength extremes is low,
and an infrared LED in the fiber robots contaminates some spectra redward of 8500\AA, with an 
amplitude that depends on the proximity of the fiber to the source.  The fibers have a diameter of 
only 1\farcs5.  We generically aimed for 30 
sky fibers, sometimes obtaining more if there was a shortage of targets, and 3--5 SDSS F-star 
candidates for flux calibration.  

In 2004 we tried to put 20 of the sky fibers on blank sky positions
selected from the SDSS imaging data for the field and the rest at random positions, but eventually
switched to simply using random positions as it became clear that contamination of the sky fibers
by sources was not a significant problem.
In the first runs in 2004 the atmospheric dispersion corrector was not working properly (see Table~\ref{tab:passes}), which means that some of the spectra
could not be properly flux calibrated and there are significant spectral distortions unless the
data was obtained very close to the zenith.  The guide cameras are primarily red sensitive, so the
fibers generally were properly positioned for the red light while the bluer wavelengths were 
systematically shifted, sometimes leading to quite dramatic losses for blue emission from point
sources.  

Observations are described by a three digit pass number ABB where, in general, A
indicates the sequential pass over the fields and BB indicates the field.  So pass 203 would
be the second pointing at sub-field 3.  The individual pointings were not exactly centered
on the fields, but were shifted to help maximize the overall completeness.  Weather problems,
leading to repeated observations, the longevity of the project, and the introduction of coadded
spectra from multiple observations eventually led to a partial break down in the naming
scheme.   Table~\ref{tab:passes} summarizes all the observations.

In 2004 we carried out three passes with integration times of 24, 45 and 75
minutes divided into 2, 3 and 4 exposures respectively.  The targets were divided
into surface brightness classes with the high, medium and low surface brightness targets
assigned to the short, medium, and long integration times.  Targets with failed redshifts
in the first passes were recycled for observations in the later, deeper passes.  The systematic
recycling of failures during this and later seasons means that fiber collisions are largely irrelevant
to the completeness of any of the AGES samples.  

The 2005 observations used pass numbers of 4 through 7 indicating the month of the
observation (March, April, May and June/July) and field numbers of $1\cdots 26$ where
$1\cdots15$ correspond to the standard sub-fields, $16\cdots 21$ to observations in the
boundary regions, and $22 \cdots 26$ to repeat observations in the 15 standard sub-fields.
The exposure times were generally 90 minutes.  Experiments using 54 minute exposure
times had poor redshift yields.
The observing conditions were not homogeneous, with significant
variations in the signal-to-noise ratios beyond the effects of the changing exposure times.

In 2006 we had less time and terrible weather.  All the observations were designated
as pass 8, where in the first run we observed fields 1--5 (801, 802, 803, 804, 805).
The poor yields led us to repeat these observations in the second run (these
we labeled by the field number plus twenty, so 821, 822, 823, 824, 825, and 834 for observations
of fields 1, 2, 3, 4, 5 and 14, and there was one additional observation of field 1 
labeled 841). We also produced co-added spectra of all multiply observed targets
that were assigned codes of 861, 862, and 863.  In 2007 we tried to focus on fields
with lower completeness levels.  These were numbered in the 900's, again adding 20
to the pass number when a field was re-observed. 

Quasars with redshift $z> 2.4$ were repeatedly reobserved until the co-added spectrum
yielded a signal-to-noise ratio above 10/pixel.  The objective was to build a 
clean sample for potentially studying correlations in Ly$\alpha$ forest absorption.
SDSS redshifts are marked as pass/aperture 0/0 entries.

\section{Data Reduction}
\label{sec:data}

The data were reduced by two separate pipelines, the standard Hectospec pipeline at the
Center for Astrophysics (CfA) and a modified SDSS pipeline, HSRED. 

In the CfA pipeline
the separate exposures were de-biased and flat fielded using exposures of the MMT ceiling
illuminated by a continuum lamp (the latter exposures had
the lamp spectral shape removed by the IRAF program ``apflatten").
The object exposures were then compared before extraction to allow identification
and elimination of cosmic rays through interpolation.  Spectra were then
extracted from individual exposures using the variance weighting method,
wavelength calibrated and combined. Each fiber has a  distinct
wavelength dependence in throughput, which can be estimated using
flat field exposures or the twilight sky. The object spectra
were next corrected for this dependence, followed by a correction
to put all the spectra on the same exposure level. The latter correction was
estimated by the strength of several night sky emission lines.   Sky subtraction
was performed, using object-free spectra as near as possible to each target.
Small corrections to the wavelength zero point based on the wavelengths of night sky emission lines 
were then applied.  Finally,
redshifts were estimated by cross correlation with emission/absorption 
line galaxy and AGN template spectra.  The CfA pipeline spectra are then
the average of the extracted spectra in counts.

For HSRED, the observations of the flat-field screen taken in the
afternoon were again used to correct for the high-frequency flat-field
variations and fringing in the CCD.  We removed low-frequency
fiber-to-fiber transmission differences using observations of the twilight
sky.   Wavelength solutions were obtained each night using observations of
HeNeAr calibration lamps, and the locations of strong emission lines in
the spectrum of the night sky were used to correct for any drift in the
wavelength solution between observations of the calibration frames and the
data frames. Each Hectospec configuration has approximately 30 fibers
dedicated to measuring the sky spectrum.  These sky observations were used
to create a median sky spectrum for each exposure which was interpolated
and subtracted from each object spectrum. Simultaneous observations of
F-type stars in each configuration were cross-correlated against a grid of
Kurucz models (Kurucz 1993) to derive a sensitivity function for each
observation, thus linking the observed counts to absolute flux units.
Where flux calibration is successful, the HSRED spectra are $F_\lambda$ 
in units of $10^{-17}{\rm\;ergs\;cm^{-2}\;s^{-1}\;\AA^{-1}}$.

Redshifts were determined using programs available in the IDLSPEC2D package
of IDL routines developed for the SDSS. To determine the redshift of each
object in the survey, we compared the observed spectra with empirical
stellar, galaxy, and quasar template models included in the IDLSPEC2D
package and allowed the strength of the emission lines present in the object
to be fit simultaneously with the redshift of the galaxy.  The final
redshift and object classification were determined by selecting the
template and redshift combination that minimized the $\chi^2$ between
model and data.

All spectra were visually inspected, usually by two individuals 
(CSK and DJE), with a particular focus on low S/N spectra and 
spectra where the two pipelines produced discrepant 
redshifts.  These were then either flagged as wrong, adjusted 
to the correct value or analyzed manually.

\begin{figure}[p]
\centerline{\includegraphics[width=5.0in]{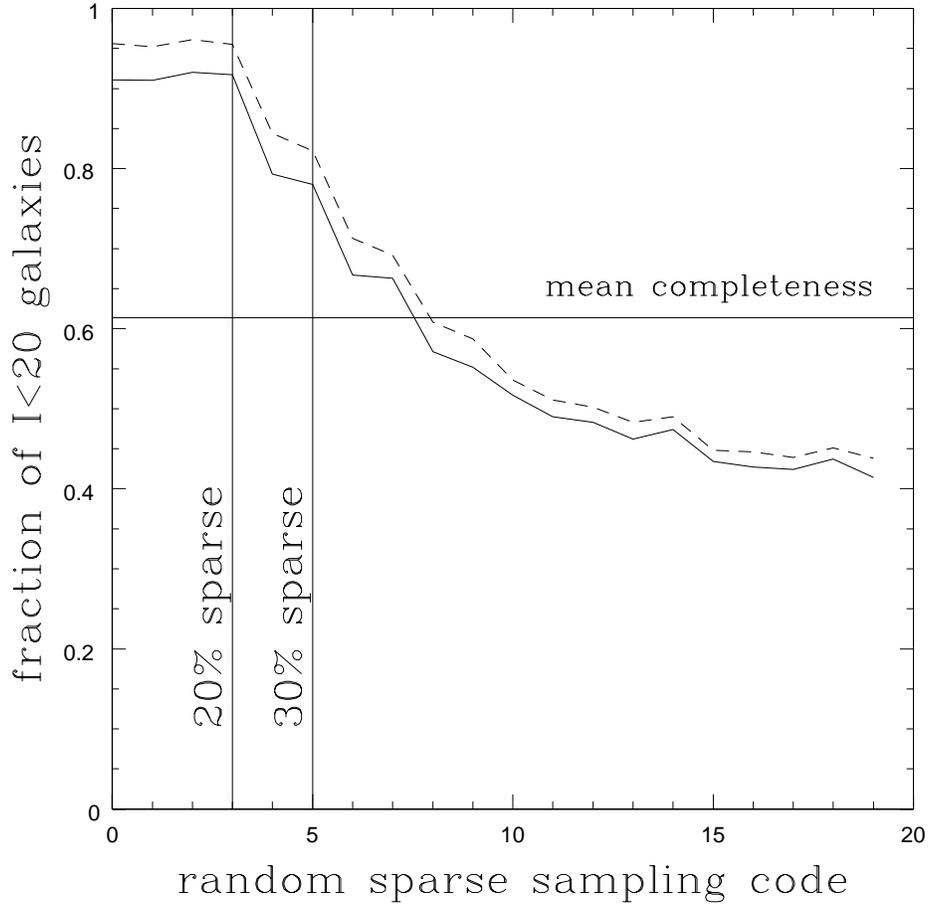}}
\caption{ Completeness as a function of the random sparse sampling {\bf rcode} (see \S2.2) 
   for all $\hbox{I}<20$~mag galaxies, where 
   each $\hbox{rcode}$ bin contains an average of 5\% of the targets
   (1959 for the $\hbox{I}<20$~mag galaxies).  The dashed
   line indicates the fraction with spectra and the solid line the
   fraction with measured redshifts.  The horizontal line shows the 
   mean completeness and the vertical lines mark the 20\% and 30\%
   sparse sampling goals for the main I band sample and the other
   galaxy samples, respectively. 
   }
\label{fig:rcode}
\end{figure}

\begin{figure}[p]
\centerline{\includegraphics[width=5.0in]{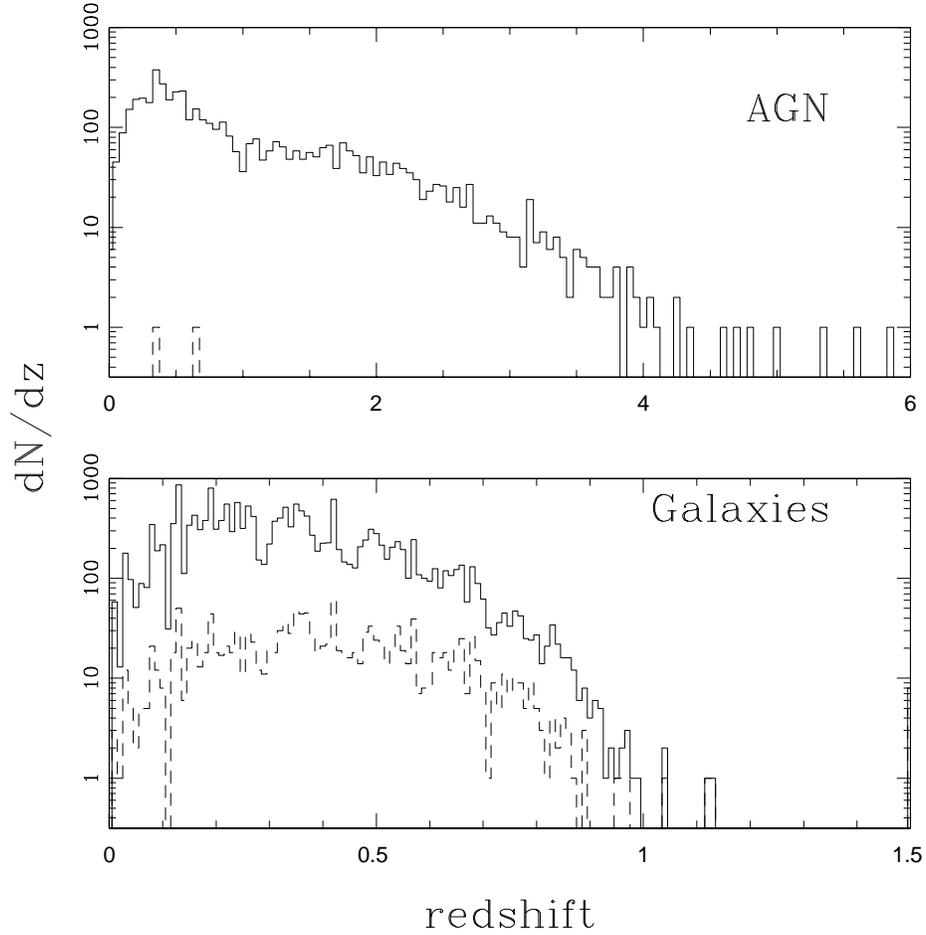}}
\caption{  The redshift distributions of AGN (top) and galaxies (bottom).  In the
  top panel the solid (dashed) histograms are for point (extended) sources with
  an AGN targeting code ($\hbox{qcode06}>3$).  In the lower panel, the solid
  histograms shows all objects targeted as galaxies ($\hbox{gcode06}>0$), and
  the dashed histograms show the objects targeted as galaxies that also had
  an AGN targeting code ($\hbox{qcode06}>3$). 
   }
\label{fig:zdist}
\end{figure}

\begin{figure}[p]
\centerline{\includegraphics[width=5.0in]{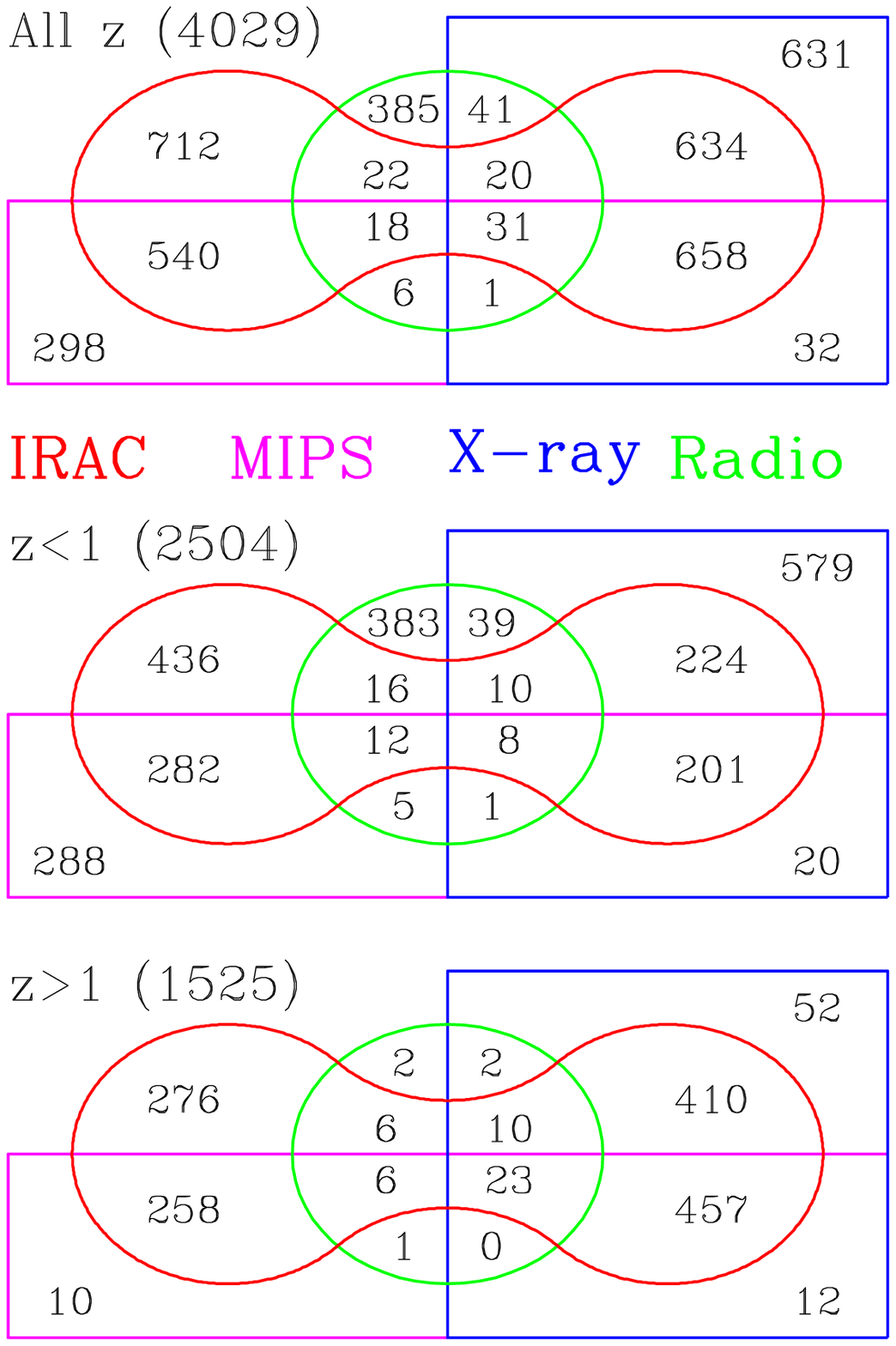}}
\caption{ 
    Venn diagrams illustrating the degree of overlap between the primary
   AGN selection methods (IRAC mid-IR, MIPS $24\mu$m, X-ray and radio).  
   The panels show the divisions for all (top panel), lower redshift 
   ($z<1$, middle panel) and higher redshift ($z>1$, lower panel) 
   AGN.  For any one of the selection methods, there are 8 possible
   overlaps ranging from no other method to all three other methods.
   }
\label{fig:venn}
\end{figure}

\section{A Summary of the Survey }
\label{sec:summary}

The general properties of the survey are summarized in Tables~\ref{tab:fields},
\ref{tab:samples1} and \ref{tab:samples2} and illustrated in Figures~\ref{fig:fields},
\ref{fig:rcode} and \ref{fig:zdist}.  
Table~\ref{tab:fields} and Figure~\ref{fig:fields}
illustrate the spatial completeness of the survey using the main I-band 
galaxy sample.  In the standard sub-fields, spectra were attempted for 96.6\%
of this sample, and redshifts were measured for 93.6\%.
The completeness is worst for fields 13, 14 and 15, both in terms of
the fraction of attempts (86\% to 91\%) and the overall completeness
(83\% to 89\%).  Two factors led to the lower completeness. First,
all three of the fields have more targets (786, 748 and 855 respectively)
than the mean (734 per field),
although we achieved much higher completenesses for other dense 
fields such as field 4.  Second, we emphasized completing the lower
field numbers in the face of poor weather and limited time to
finish our observations.  Every field was observed many times
(see Table~\ref{tab:passes}), so fiber collisions play a very small 
role in the incompleteness.

Table~\ref{tab:samples1} summarizes the 2006/2007
samples, excluding the flux calibration stars, brown dwarf and optical
quasar test samples.  The well-defined galaxy samples
are very complete, with the main
I-band sample having the lowest completeness (94\%), followed by
the MIPS sample (95\%).  The remainder have completenesses above
98\%.  The GALEX samples are spatially inhomogeneous and 
of limited use.  In total, we obtained redshifts for roughly 61\% of the 
galaxies with $\hbox{I}<20$~mag.  

Fig.~\ref{fig:rcode}  shows the completeness
as a function of the random sample code.  Because we emphasized
observing lower rcodes, it is relatively easy to rapidly increase
the size of the sample with high completeness. 
The main I-band galaxy sample used 20\% sparse sampling
($\hbox{rcode}\leq 3$)
for its fainter magnitudes, $18.5 \leq \hbox{I} \leq 20$, while
the IR samples used 30\% sparse sampling 
($\hbox{rcode}\leq 5$).  The higher $\hbox{rcode}$s were assigned
priorities that dropped with every increase in the $\hbox{rcode}$
by two, leading to the steady drop in the completeness.  With this
design, little effort is needed to produce significantly larger
complete samples.  About 500 redshifts are needed to complete the
main galaxy sample.  Another 1600 would complete the sample to a
sparse sampling fraction between $18.5 < \hbox{I} < 20$ of 40\%. 

In total we selected almost 8977 objects as AGN candidates, 
took spectra of 7102, and obtained redshifts for 5217 of 
which 4764 were not Galactic stars.  Table~\ref{tab:samples2}
summarizes the completeness of the various categories of AGN,
breaking the statistics into the various sub-samples (point
source, bright extended sources, faint extended sources) and
giving statistics for all the AGN selection methods
(all objects with an AGN selection code) as compared to the 
total galaxy sample (all objects with $\hbox{code06} \geq 128$).
In total, we identified $1718$ AGN with $z>1$ in the field, a
surface density of more than $200$/deg$^2$.  Three quasars with
redshifts above 5 were identified (\citealt{Cool2006}). 
A redshift $6.12$ quasar was targeted as an IRAC AGN but
not observed before it was discovered by \cite{McGreer2006} and 
\cite{Stern2007}.  The completenesses for
the point source and bright extended AGN are generally good,
while that for the fainter extended AGN candidates is very 
poor.  Fig.~\ref{fig:zdist} shows the redshift distributions
of the galaxy and AGN populations.  

The different AGN selection
methods emphasize different galaxy types and redshift ranges,
as discussed in more detail by \cite{Hickox2007}, \cite{Gorjian2008},
\cite{Assef2010} and \cite{Assef2011}.  Fig.~\ref{fig:venn}
illustrates some of these issues using a Venn diagram adapted
from \cite{Assef2010} showing the overlap between the WSRT, 
X-ray, IRAC and MIPS quasar selection methods for several different 
redshift ranges.  The primary difference between the X-ray
(and radio) sample versus the IRAC and MIPS samples is that
X-ray selection is essentially independent of host properties
while the IRAC and MIPS samples are not.  Thus, lower 
redshift AGN are more likely to be X-ray selected because
the generally larger contribution of the host galaxy at
lower redshifts changes the mid-IR colors or makes the
optical counterpart to the MIPS source non-point-like.
On the other hand, the mid-IR selection methods may well
be better for finding moderately obscured quasars where
the soft X-ray photons to which Chandra is most sensitive 
are absorbed.  Many of these problems could be solved using the \cite{Assef2010}
template models to fit the complete photometry for each
source and target those with any evidence of an AGN
contribution. 

\begin{figure}[p]
\centerline{\includegraphics[width=6.0in]{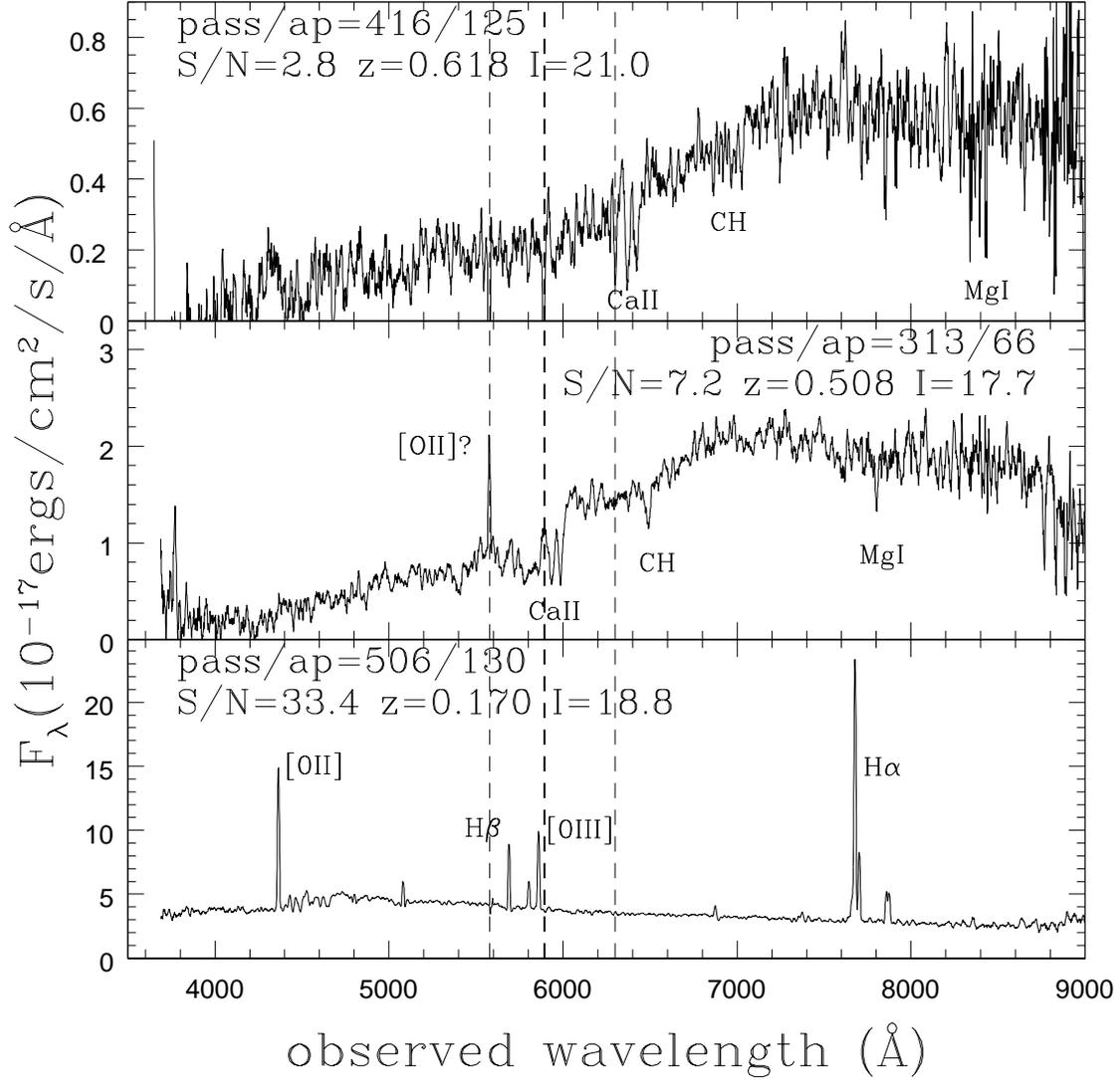}}
\caption{ 
   Three examples of the spectra of galaxies.  The top, middle and lower 
   panels show spectra with continuum signal-to-noise ratios typical of 
   the worst 5\%, median and best 5\% of the spectra yielding redshifts.
   Several features are labeled, and the vertical lines mark the strong
   sky lines.  The spectra are smoothed by an 11 pixel box car, which
   roughly halves the intrinsic spectral resolution.
   }
\label{fig:spec1}
\end{figure}

\begin{figure}[p]
\centerline{\includegraphics[width=6.0in]{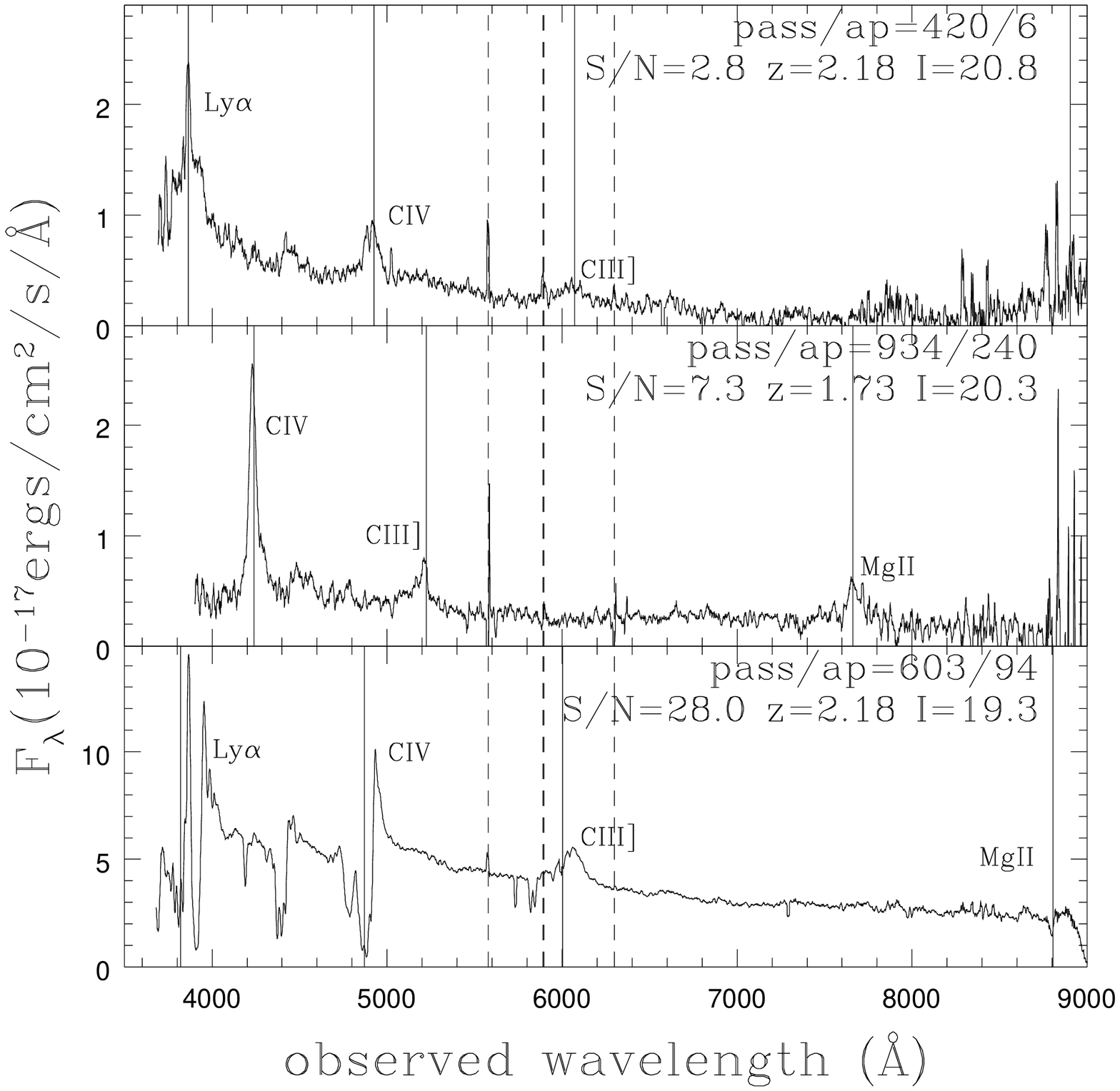}}
\caption{ 
   Three examples of the spectra of quasars.  The top, middle and lower  
   panels show spectra with continuum signal-to-noise ratios typical of 
   the worst 5\%, median and best 5\% of the spectra yielding redshifts.
   Several features are labeled and marked by the solid vertical lines
   while the dashed lines mark the strongest
   sky lines.  The spectra are smoothed by an 11 pixel box car. The
   strong (BAL) absorption features in the bottom spectrum have 
   biased the pipeline redshift estimate.
   }
\label{fig:spec2}
\end{figure}

\section{Data Release}
\label{sec:release}

The AGES data release consists of a series of row-matched tables. They do not
contain all entries from the matched photometric catalogs, as these contain far more
information than is needed to interpret the AGES data.  We have also
included only the final photometry used in later observing seasons.  
We do not report the photometry associated with the intermediate states because
the completeness of the final samples is so high.  We do report the
selection codes used in the earlier observing seasons so that the
history of the targeting can be traced if necessary and to explain
the origin of the small numbers sources with redshifts that were not
explicitly targeted in the final seasons.

Table~\ref{tab:codes} summarizes the selection codes for each source.  This
includes the 2004, 2005 and 2006/2007 targeting codes, the field IDs, the
random sparse sampling code, the bright star flag, the IRAC AGN sub-sample
code, and the flags for whether the source was considered a standard galaxy,
a quasar, a point source, or a fainter galaxy that was an AGN candidate. 

Table~\ref{tab:mags} presents the photometry for the sources from XBo\"otes
(\citealt{Murray2005}, \citealt{Kenter2005}, \citealt{Brand2006}),
GALEX (\citealt{Martin2005}), NDWFS DR3 (\citealt{Jannuzi1999}), zBo\"otes (\citealt{Cool2007}),
FLAMEX (\citealt{Elston2006}), the IRAC Shallow Survey (\citealt{Eisenhardt2004}) 
and \cite{Soifer2004}.  The X-ray photometry is in counts for sources with
a 25\% or greater Bayesian match probability in \cite{Brand2006}.  The
NDWFS, FLAMEX and IRAC Shallow Survey are the (Kron-like/mag$\_$auto) Vega magnitudes,
GALEX and zBo\"otes are in AB magnitudes, and the $24\mu$m flux is in mJy.
Objects with unusual photometric properties should be inspected closely
before use.  Table~\ref{tab:redshift} summarizes the spectroscopy, listing the number
of spectra taken, the estimated redshifts, the (continuum) signal-to-noise ratio of 
the spectra, and the pass/aperture identification code for each 
spectrum.    Figure~\ref{fig:spec1} shows three examples
of spectra of galaxies illustrating the quality for (continuum)
signal-noise-ratios representative of the worst 5\%, median and best 5\% of
the sample.  Figure~\ref{fig:spec2} does the same for three
examples of quasar spectra.  The signal-to-noise estimate in
Table~\ref{tab:redshift} is an indicator of redshift reliability,
as well as any agreements/disagreements between repeated low signal-to-noise 
spectra.

Table~\ref{tab:photoz} provides photometric redshift estimates and 
template decompositions for each object's photometry following 
\cite{Assef2010}.  The SED is fit as a combination of an early-type,
late-type, star forming and (obscured) AGN for the full range of
available UV/optical/near-IR/far-IR photometry, and we report
the luminosities associated with each component and the extinction
applied to the AGN template.  Where a spectroscopic redshift is
available, the template decomposition is carried out at the 
spectroscopic redshift.  The accuracy of the photometric 
redshifts is $\sigma_z=0.04/(1+z)$ for galaxies, and \cite{Assef2010}
should be consulted for a detailed discussion of the results
for strong AGN.   Finally, Table~\ref{tab:spectra} presents the
complete set of spectra in whatever form was available.  This
is somewhat heterogeneous in terms of pipeline and flux calibration,
but a complete, homogeneous re-reduction of the data is beyond the
scope of this paper. 

As discussed in \S\ref{sec:data}, the NDWFS Sextractor Kron-like magnitudes $I_{AUTO}$ 
tend to overestimate source fluxes near bright stars.  In \cite{Cool2011},
we developed a method to produce a corrected estimate, which we summarize
here.  Let $I_R = R_{AUTO} + (I(6\farcs0)-R(6\farcs0)))$ be the I-band 
magnitude predicted from the R-band Kron-like magnitude $R_{AUTO}$ and 
the $6\farcs0$ aperture color.  The surrogate I-band total magnitude
\begin{equation}
    I_{tot} = { I_{AUTO} + I_R \over 2 } f + (1-f) \hbox{max}(I_{AUTO},I_R)
\end{equation}
where $f = \exp(-(I_{AUTO}-I_R)^2/0.2^2)$ is a weight factor.  On
average $\langle I_{tot}-I_{AUTO}\rangle = 0.005$~mag with an rms
scatter of 0.02 mag, but 10\% (5\%) of galaxies have shifts of 0.1 (0.5)~mag.

Completeness corrections for the galaxy samples are relatively straightforward.
\cite{Cool2011} discusses several tests of the completeness of the input
catalogs, coming to the conclusion that the catalog completeness is of
order 96-97\%, largely due to the loss of faint objects superposed on brighter 
galaxies or stars. There are very few spurious objects.  Only 1\% of the main-sample 
galaxies lack counterparts in SDSS imaging, although some saturated stars 
were mis-identified as galaxy targets.  These are easily identified because 
the resulting spectra and redshifts are stellar.  The remaining issues
are the sparse sampling fractions, fiber allocation completeness and
redshift failure rates.  Following \cite{Cool2011}, Table~\ref{tab:stats}
provides the completeness corrections for the galaxy samples as well as
the maximum redshift at which the galaxy would have entered the AGES
sample and the corresponding volume $V_{max}$ in the survey.

\section{Discussion}

In summary, the AGN and Galaxy Evolution Survey has measured approximately
23745 redshifts
in the Bo\"otes field of the NDWFS using a layered approach to target selection
that produced well-defined samples of galaxies and AGN over a broad wavelength
range.  Here we have outlined the selection functions used during the survey,
summarized the general properties of the resulting samples, and released the
redshift data with a sketch of the underlying photometry.  For the full set
of photometric data, users must consult the original surveys.  

AGES contains well-defined, highly complete galaxy samples in the optical
B$_W$, R and I bands, the near-IR J and K/K$_s$ bands, and the mid-IR
IRAC and MIPS $24\mu$m bands.  These have been used to derive luminosity
functions and their evolution in the optical (\citealt{Cool2011}), all four
IRAC bands (\citealt{Dai2009}) and at $24\mu$m (\citealt{Huang2007}, \citealt{Rujopakarn2010}).
The extensive redshift information can then be used to calibrate and test photometric
redshifts (\citealt{Brodwin2006}, \citealt{Brown2007}, 
\citealt{Assef2008}, \citealt{Assef2010}, \citealt{Hildebrandt2010}) 
that can then be used to search for high redshift clusters in the field 
(\citealt{Eisenhardt2008}).  
Combining the broad range of source types, extensive photometry and large
number of redshifts \cite{Assef2008} and \cite{Assef2010} built SED 
template models covering the range $0.1\mu$m to $24\mu$m for both 
galaxies and quasars.  The set of four templates can describe almost
all the sources in the sample well, and can be easily adapted to
other filter systems.

The AGES redshift data has also been used to help estimate
bolometric corrections from the mid to far-IR (\citealt{Bavouzet2008}),
where there have been significant questions about how to correct from
$24\mu$m fluxes to total far-IR fluxes.  \cite{Watson2009} used it to
estimate the X-ray properties of otherwise undetected galaxies and AGN,
using ``stacking'' to estimate the contribution of AGN and star formation 
to X-ray emission as a function of cosmic epoch.  \cite{Brand2009} 
used it to explore the origin of $24\mu$m emission in otherwise early-type 
galaxies, and \cite{Atlee2009} used it to study the evolution of the 
UV upturn in early-type galaxies.

The initial AGES data were used to develop a remarkably successful mid-IR
approach to quasar selection by \cite{Stern2005}.  This approach was then
used in the later years not only to build the largest existing sample of 
mid-IR-selected AGN, but also to explore its properties and limitations in
detail both through other AGES mid-IR target samples (\citealt{Assef2010},
\citealt{Assef2011}) and comparisons with X-ray sources (\citealt{Gorjian2008},
\citealt{Hickox2007}, \citealt{Assef2011}).  \cite{Hickox2007}, \cite{Hickox2009}
and \cite{Starikova2010} use the AGES data to explore the relationships 
between AGN accretion and galaxy properties and clustering, while \cite{Kollmeier2006}
examined the Eddington ratio distribution of quasars to find that the 
distributions were surprisingly narrow.  \cite{Brown2006} and \cite{Assef2011}
examine quasar luminosity functions using mid-IR and X-ray selected samples.

AGES was also used to help design aspects of SDSS-III (\citealt{Eisenstein2011}).
Finally, the existence of the extensive AGES data has also helped motivate further
studies of the Bo\"otes field.  The Spitzer Deep, Wide-Field Survey (SDWFS, 
\citealt{Ashby2009}) doubled the depth of the original IRAC Shallow Survey
(\citealt{Eisenhardt2004}), while simultaneously enabling the first large
scale extragalactic study of the mid-IR variability of AGN (\citealt{Kozlowski2010a})
and the serendipitous discovery of a highly luminous but obscured supernova
(\citealt{Kozlowski2010b}).  In the MIPS AGN and Galaxy Evolution Survey
(MAGES, Jannuzi et al. 2011, in prep), the MIPS $24$, $70$ and $160\mu$m
data for the field were similarly improved.  The field has also been imaged
by Herschel as a GTO program.

\begin{acknowledgements}
We thank the Hectospec instrument team and all the MMT Hectospec queue 
observers for making this project possible.  We also thank
T. Soifer, D. Weedman, J. Houck, M. Rieke and collaborators for permission to use 
the results of their GTO Spitzer/MIPS survey of the Bo\"otes field.
RJA is supported by an appointment to the NASA Postdoctoral Program a the
Jet Propulsion Laboratory, administered by Oak Ridge Associated Universities
through a contract with NASA.  BTJ and AD are supported by the NSF through
its funding of NOAO, which is operated for the NSF by AURA under a
cooperative agreement.
CJ, SSM and WRF acknowledge support from the Smithsonian Institution and by NASA contracts 
NAS8-38248, NAS8-01130, NAS8- 39073, and NAS8-03060, and NASA grant GO3-4176A.
Observations reported here were obtained at the MMT Observatory, a joint facility
of the Smithsonian Institution and the University of Arizona.  This work made
use of images and/or data products provided by the NOAO Deep Wide-Field Survey,
which is supported by the National Optical Astronomy Observatory (NOAO).  NOAO
is operated by AURA, Inc., under a a cooperative agreement with the National
Science Foundation.  This work is based in part on observations made with the 
Spitzer Space Telescope, which is operated by the Jet Propulsion Laboratory, 
California Institute of Technology under a contract with NASA. Support for 
this work was provided by NASA through an award issued by JPL/Caltech.
\end{acknowledgements}

\facility{MMT, Spitzer, Chandra, GALEX, VLA, Mayall}


\begin{deluxetable}{rrrcccc}
\tablecaption{The Standard Fields}
\tablewidth{0pt}
\tablehead{ Field &\multicolumn{1}{c}{RA} &\multicolumn{1}{c}{Dec} &\multicolumn{4}{c}{Main Galaxy Sample} \\ 
                  &   &    &Sample &Spectra &Redshifts &Completeness 
  }
\startdata
 1    &216.750000     &35.365000  &774 &772 &753 &97.3\%\\
 2    &216.666667     &34.578889  &751 &750 &731 &97.3\%\\
 3    &216.766667     &33.838333  &729 &723 &710 &97.4\%\\
 4    &216.629167     &33.121389  &911 &902 &861 &94.5\%\\
 5    &217.404167     &35.402500  &688 &667 &633 &92.0\%\\
 6    &217.416667     &34.591389  &574 &572 &566 &98.6\%\\
 7    &217.441667     &33.990833  &551 &546 &529 &96.0\%\\
 8    &217.454167     &33.283889  &865 &861 &839 &97.0\%\\
 9    &218.245833     &35.326667  &652 &631 &602 &92.3\%\\
10    &218.133333     &34.712222  &728 &720 &686 &94.2\%\\
11    &218.225000     &33.922500  &614 &612 &603 &98.2\%\\
12    &218.395833     &33.411389  &785 &766 &749 &95.4\%\\
13    &219.020833     &35.464167  &786 &717 &697 &88.7\%\\
14    &218.895833     &34.618056  &748 &664 &636 &88.8\%\\
15    &219.091667     &33.860833  &855 &737 &711 &83.2\%\\
\enddata
\label{tab:fields}
\tablecomments{These are the RA/Dec of the 15 standard sub-field centers, followed by the
number of Main I-band ($\hbox{gcode06}=524288$) galaxies in the field, the number
for which spectra were obtained, the number of successful redshift measurements
and the resulting completeness.  An object is in a field if it is closer to the
center than $R_{fld}=0.49$~deg.  Objects in overlapping fields are assigned to the closest field
center, and objects in none of these standard sub-fields are given field number $-1$. See
Fig.~\ref{fig:fields} for the positions of the fields on the sky.}
\end{deluxetable}

\begin{deluxetable}{rrrrrrrrrrr}
\tablecaption{The Spectroscopic Observations}
\tablewidth{0pt}
\tablehead{ Pass &Field &\multicolumn{1}{c}{Date} &\multicolumn{1}{c}{RA} &\multicolumn{1}{c}{Dec}  &$N_{exp}$  &$T_{exp}$ &$N_{spec}$ &Air  &Mean &Comments \\
                 &      &     &   &    &           &(sec)     &           &Mass &SNR  & \\ }
\startdata
  0 &  0 &\multicolumn{5}{c}{Spectra from the SDSS Survey} &2946 & &  & \\
101 &  1 &2004.0415 &14:26:49.36 &35:22:05.57 & 4 &  2400 & 268 & 1.03 &16.21 & \\
102 &  2 &2004.0421 &14:26:58.80 &34:38:07.68 & 2 &  1800 & 266 & 1.01 &11.21 & \\
103 &  3 &2004.0416 &14:26:33.20 &33:59:51.36 & 3 &  2700 & 263 & 1.02 &16.51 & \\
104 &  4 &2004.0416 &14:26:29.20 &33:09:53.04 & 4 &  3600 & 267 & 1.19 &14.92 &major ADC \\
105 &  5 &2004.0420 &14:29:47.89 &35:28:48.00 & 2 &  1440 & 262 & 1.12 & 7.09 & \\
106 &  6 &2004.0414 &14:31:49.76 &34:50:47.40 & 3 &  2700 & 267 & 1.18 & 7.66 &not fluxed \\
107 &  7 &2004.0420 &14:29:41.89 &33:53:09.36 & 2 &  1440 & 260 & 1.03 &10.08 & \\
108 &  8 &2004.0422 &14:29:37.09 &33:13:53.04 & 2 &  1440 & 264 & 1.00 &11.02 & \\
109 &  9 &2004.0420 &14:32:38.18 &35:23:05.99 & 2 &  1440 & 259 & 1.06 & 7.65 & \\
110 & 10 &2004.0414 &14:31:49.76 &34:50:47.40 & 3 &  2700 & 267 & 1.04 & 8.98 &not fluxed \\
111 & 11 &2004.0416 &14:32:40.98 &33:53:51.37 & 3 &  2700 & 263 & 1.02 &18.17 & \\
112 & 12 &2004.0420 &14:33:35.78 &33:22:35.04 & 2 &  1440 & 268 & 1.01 &10.06 & \\
113 & 13 &2004.0416 &14:35:32.87 &35:24:48.00 & 3 &  2700 & 258 & 1.20 &14.22 &major ADC \\
114 & 14 &2004.0421 &14:35:36.87 &34:39:37.68 & 2 &  1440 & 270 & 1.01 & 8.53 & \\
115 & 15 &2004.0420 &14:35:36.87 &33:58:51.36 & 2 &  1440 & 270 & 1.09 & 9.74 & \\
201 &  1 &2004.0421 &14:26:26.80 &35:22:36.00 & 3 &  2700 & 259 & 1.09 & 9.00 & \\
202 &  2 &2004.0422 &14:26:58.80 &34:38:07.68 & 3 &  2700 & 258 & 1.07 & 8.00 & \\
203 &  3 &2004.0421 &14:26:33.20 &33:59:51.36 & 3 &  2700 & 259 & 1.28 & 9.03 &major ADC \\
204 &  4 &2004.0421 &14:26:43.20 &33:08:47.04 & 3 &  2700 & 268 & 1.04 & 8.87 & \\
205 &  5 &2004.0422 &14:29:25.09 &35:20:36.00 & 3 &  2700 & 244 & 1.29 & 6.94 &major ADC \\
206 &  6 &2004.0422 &14:29:38.29 &34:42:01.68 & 3 &  2700 & 262 & 1.10 &10.89 & \\
207 &  7 &2004.0611 &14:29:46.29 &33:59:27.34 & 4 &  4500 & 264 & 1.08 &10.21 & \\
208 &  8 &2004.0423 &14:29:37.09 &33:13:53.04 & 2 &  1800 & 264 & 1.32 & 5.60 &major ADC \\
209 &  9 &2004.0612 &14:32:59.18 &35:19:36.00 & 4 &  4500 & 276 & 1.08 &10.87 &not fluxed \\
210 & 10 &2004.0422 &14:32:39.38 &34:40:49.68 & 3 &  2700 & 257 & 1.01 & 9.57 & \\
211 & 11 &2004.0423 &14:32:42.98 &33:54:21.36 & 3 &  2220 & 260 & 1.01 & 6.54 & \\
212 & 12 &2004.0423 &14:33:35.78 &33:22:35.04 & 3 &  2700 & 261 & 1.06 & 6.60 & \\
213 & 13 &2004.0420 &14:35:45.27 &35:28:48.00 & 3 &  2700 & 261 &-1.00 & 9.34 &major ADC \\
214 & 14 &2004.0423 &14:35:36.87 &34:39:37.68 & 3 &  2700 & 254 & 1.11 & 5.74 & \\
215 & 15 &2004.0423 &14:35:36.87 &33:58:51.36 & 3 &  2700 & 259 & 1.01 & 7.20 & \\
301 &  1 &2004.0610 &14:27:00.40 &35:21:54.01 & 4 &  4500 & 279 & 1.18 &10.70 &major ADC \\
302 &  2 &2004.0620 &14:26:40.40 &34:34:43.68 & 4 &  4500 & 262 & 1.30 & 9.61 &ADC off \\
303 &  3 &2004.0622 &14:27:04.00 &33:50:18.36 & 4 &  4500 & 268 &-1.00 &10.93 & \\
304 &  4 &2004.0621 &14:26:31.00 &33:07:17.04 & 4 &  4500 & 264 & 1.37 & 9.68 &major ADC \\
305 &  5 &2004.0616 &14:29:36.69 &35:24:09.00 & 4 &  4500 & 272 & 1.06 & 9.33 & \\
306 &  6 &2004.0615 &14:29:40.09 &34:35:28.70 & 4 &  4500 & 265 & 1.06 &12.24 & \\
307 &  7 &2004.0615 &14:29:46.29 &33:59:27.40 & 4 &  4500 & 263 & 1.30 & 9.53 &major ADC \\
308 &  8 &2004.0621 &14:29:49.49 &33:17:02.04 & 4 &  4500 & 272 & 1.07 &10.92 & \\
309 &  9 &2004.0616 &14:32:59.18 &35:19:36.00 & 4 &  4500 & 268 & 1.29 & 9.45 &major ADC \\
310 & 10 &2004.0613 &14:32:32.37 &34:42:43.69 & 4 &  4500 & 265 & 1.04 &11.61 &not fluxed \\
311 & 11 &2004.0614 &14:32:53.97 &33:55:21.36 & 4 &  4500 & 266 & 1.07 & 9.04 &not fluxed \\
312 & 12 &2004.0626 &14:33:35.18 &33:24:41.04 & 5 &  5625 & 260 & 1.03 & 6.59 &scattered light \\
313 & 13 &2004.0617 &14:36:04.86 &35:27:50.99 & 4 &  4500 & 266 &-1.00 & 8.95 &ADC off \\
314 & 14 &2004.0618 &14:35:34.87 &34:37:04.68 & 4 &  4500 & 270 &-1.00 &10.94 &ADC off \\
315 & 15 &2004.0619 &14:36:21.86 &33:51:39.35 & 4 &  4500 & 267 & 1.15 &10.96 &ADC off \\
401 &  1 &2005.0312 & 14:26:42.00 & 35:26:39.00 & 5 &  5400 & 261 & 1.05 & 9.12 & \\
402 &  2 &2005.0314 & 14:26:29.99 & 34:35:59.00 & 4 &  4080 & 241 & 1.03 & 3.82 & \\
403 &  3 &2005.0310 & 14:26:54.00 & 33:53:18.00 & 5 &  5100 & 261 & 1.04 &11.17 & \\
404 &  4 &2005.0311 & 14:26:08.00 & 33:10:02.00 & 5 &  5400 & 253 & 1.04 &18.58 & \\
405 &  5 &2005.0315 & 14:29:25.00 & 35:28:54.00 & 2 &  1800 & 253 & 1.03 & 3.09 & \\
406 &  6 &2005.0317 & 14:29:30.00 & 34:36:13.99 & 5 &  5400 & 250 & 1.05 &13.39 & \\
407 &  7 &2005.0317 & 14:29:25.00 & 33:59:56.99 & 5 &  5400 & 249 & 1.33 & 5.49 & \\
408 &  8 &2005.0317 & 14:29:33.60 & 33:21:38.00 & 6 &  6480 & 247 & 1.04 &11.42 & \\
409 &  9 &2005.0318 & 14:32:57.60 & 35:25:27.02 & 1 &  1080 & 244 & 1.00 & 6.38 & \\
410 & 10 &2005.0316 & 14:32:31.00 & 34:42:44.00 & 5 &  5400 & 253 & 1.04 &10.37 & \\
411 & 11 &2005.0316 & 14:33:01.39 & 33:59:08.99 & 6 &  6480 & 246 & 1.04 &10.27 & \\
416 &  4 &2005.0308 & 14:26:23.99 & 32:53:20.00 & 4 &  4320 & 260 & 1.03 &15.12 & \\
417 &  8 &2005.0308 & 14:29:58.00 & 33:00:17.00 & 6 &  6480 & 261 & 1.08 &18.75 & \\
418 & 12 &2005.0311 & 14:33:59.00 & 33:15:41.01 & 6 &  5400 & 259 & 1.30 &11.45 & \\
419 & 13 &2005.0311 & 14:36:43.00 & 35:25:00.00 & 4 &  4320 & 258 & 1.04 &13.25 & \\
420 & 14 &2005.0310 & 14:36:19.20 & 34:32:01.99 & 5 &  4800 & 253 & 1.03 &14.06 & \\
421 & 15 &2005.0312 & 14:36:52.20 & 33:54:17.99 & 5 &  5400 & 261 & 1.03 &17.77 & \\
422 &  1 &2005.0309 & 14:27:06.00 & 35:23:23.90 & 5 &  5400 & 253 & 1.03 & 8.46 & \\
423 &  4 &2005.0314 & 14:26:20.99 & 33:07:01.90 & 5 &  5400 & 252 & 1.04 & 6.69 & \\
501 &  1 &2005.0406 & 14:26:48.34 & 35:25:45.65 & 5 &  4500 & 252 & 1.03 & 8.66 & \\
502 &  2 &2005.0406 & 14:26:39.39 & 34:34:54.39 & 3 &  3300 & 264 & 1.19 & 5.99 & \\
503 &  3 &2005.0409 & 14:26:38.90 & 33:43:36.41 & 4 &  3640 & 262 & 1.03 & 6.09 & \\
504 &  4 &2005.0408 & 14:26:22.44 & 33:07:01.47 & 5 &  5400 & 262 & 1.34 & 4.27 & \\
505 &  5 &2005.0409 & 14:30:35.16 & 35:30:22.56 & 4 &  4320 & 258 & 1.16 & 4.50 & \\
506 &  6 &2005.0407 & 14:30:46.60 & 34:51:20.84 & 5 &  5400 & 260 & 1.06 & 8.86 & \\
507 &  7 &2005.0410 & 14:30:18.65 & 34:00:55.58 & 4 &  4320 & 258 & 1.03 & 6.70 & \\
508 &  8 &2005.0410 & 14:30:46.73 & 33:12:25.94 & 5 &  4500 & 257 & 1.20 &10.96 & \\
510 & 10 &2005.0411 & 14:32:07.71 & 34:40:38.06 & 5 &  5400 & 258 & 1.06 & 9.17 & \\
511 & 11 &2005.0411 & 14:33:05.26 & 34:00:44.66 & 4 &  4320 & 261 & 1.02 & 7.05 & \\
512 & 12 &2005.0411 & 14:33:33.74 & 33:23:28.63 & 4 &  4320 & 254 & 1.17 & 5.74 & \\
522 &  1 &2005.0405 & 14:26:19.24 & 35:12:59.23 & 3 &  3240 & 215 & 1.06 & 3.96 & \\
523 &  2 &2005.0405 & 14:26:28.60 & 34:35:57.20 & 3 &  3240 & 238 & 1.21 & 1.49 & \\
524 &  3 &2005.0406 & 14:26:49.00 & 33:49:38.72 & 3 &  3240 & 232 & 1.19 & 5.49 & \\
525 &  7 &2005.0407 & 14:29:12.19 & 34:16:33.70 & 2 &  1920 & 233 & 1.21 & 3.43 & \\
526 &  2 &2005.0406 & 14:26:49.79 & 34:48:52.29 & 4 &  4320 & 226 & 1.03 & 8.46 & \\
601 &  1 &2005.0506 & 14:26:46.00 & 35:26:22.00 & 5 &  5400 & 256 & 1.05 & 4.65 & \\
602 &  2 &2005.0507 & 14:26:32.00 & 34:36:44.00 & 6 &  6300 & 263 & 1.07 & 6.58 & \\
603 &  3 &2005.0510 & 14:26:50.00 & 33:53:03.00 & 4 &  4320 & 253 & 1.02 & 7.75 & \\
604 &  4 &2005.0510 & 14:26:22.00 & 33:07:01.99 & 2 &  2160 & 256 & 1.01 & 5.04 & \\
605 &  5 &2005.0509 & 14:29:25.00 & 35:28:54.00 & 5 &  5220 & 261 & 1.06 & 7.27 & \\
606 &  6 &2005.0508 & 14:29:30.00 & 34:36:13.98 & 6 &  6480 & 259 & 1.35 & 6.63 & \\
607 &  7 &2005.0509 & 14:29:25.00 & 33:59:56.99 & 5 &  5400 & 258 & 1.36 & 6.91 & \\
608 &  8 &2005.0510 & 14:29:57.60 & 33:16:10.01 & 4 &  4320 & 262 & 1.16 &10.97 & \\
609 &  9 &2005.0512 & 14:32:57.59 & 35:25:27.02 & 5 &  5400 & 261 & 1.06 &11.52 & \\
610 & 10 &2005.0510 & 14:32:31.00 & 34:42:44.00 & 3 &  3240 & 254 & 1.51 & 3.70 & \\
611 & 11 &2005.0511 & 14:33:01.39 & 33:59:08.99 & 4 &  4320 & 258 & 1.10 &10.07 & \\
612 & 12 &2005.0514 & 14:33:55.00 & 33:22:44.99 & 5 &  5400 & 256 & 1.05 &12.57 & \\
613 & 13 &2005.0514 & 14:36:27.00 & 35:28:20.00 & 6 &  6480 & 261 & 1.37 &14.64 & \\
622 &  5 &2005.0511 & 14:29:25.00 & 35:28:54.00 & 5 &  5400 & 258 & 1.47 & 5.85 & \\
709 &  9 &2005.0705 & 14:33:11.90 & 35:20:39.00 & 5 &  5400 & 257 & 1.23 & 7.34 & \\
710 & 10 &2005.0703 & 14:33:40.35 & 34:31:42.01 & 6 &  6480 & 259 & 1.09 & 3.76 & \\
712 & 12 &2005.0702 & 14:33:18.70 & 33:24:50.99 & 5 &  5400 & 260 & 1.05 & 9.78 & \\
713 & 13 &2005.0703 & 14:36:09.99 & 35:26:29.99 & 5 &  5400 & 258 & 1.50 & 8.96 & \\
714 & 14 &2005.0630 & 14:35:36.37 & 34:36:27.99 & 5 &  5400 & 258 & 1.12 & 8.22 & \\
715 & 15 &2005.0701 & 14:36:25.15 & 33:52:09.97 & 5 &  5400 & 258 & 1.08 &13.82 & \\
722 & 15 &2005.0706 & 14:36:35.40 & 33:44:30.00 & 5 &  5400 & 241 & 1.14 &11.81 & \\
801 &  1 &2006.0324 & 14:26:41.60 & 35:26:38.18 & 6 &  7200 & 256 & 1.12 & 7.34 & \\
802 &  2 &2006.0326 & 14:26:29.99 & 34:35:59.00 & 7 &  8400 & 263 & 1.14 & 8.07 & \\
803 &  3 &2006.0326 & 14:26:54.00 & 33:53:17.99 & 7 &  8400 & 265 & 1.09 &10.16 & \\
804 &  4 &2006.0331 & 14:26:29.56 & 33:07:10.40 & 7 &  8400 & 266 & 1.07 & 8.08 & \\
805 &  5 &2006.0404 & 14:29:56.92 & 35:27:17.80 & 5 &  5107 & 261 & 1.04 & 6.08 & \\
821 &  1 &2006.0427 & 14:26:41.61 & 35:26:38.20 & 3 &  3600 & 252 & 1.14 & 5.18 & \\
822 &  2 &2006.0429 & 14:26:30.00 & 34:35:59.00 & 6 &  6600 & 260 & 1.33 & 7.69 & \\
823 &  3 &2006.0430 & 14:26:54.00 & 33:53:18.00 & 5 &  6000 & 267 & 1.04 & 9.01 & \\
824 &  4 &2006.0430 & 14:26:29.56 & 33:07:10.40 & 5 &  6000 & 261 & 1.29 &10.56 & \\
825 & 14 &2006.0501 & 14:35:52.01 & 34:37:31.81 & 6 &  7200 & 263 & 1.34 & 6.04 & \\
834 &  5 &2006.0501 & 14:29:55.77 & 35:27:17.81 & 5 &  6000 & 262 & 1.04 & 9.77 & \\
841 &  1 &2006.0429 & 14:26:41.61 & 35:26:38.18 & 5 &  6000 & 251 & 1.04 & 7.88 & \\
861 & \multicolumn{6}{c}{Coadded Spectra}                   & 700 &      &14.24 & \\
862 &  \multicolumn{6}{c}{Coadded Spectra}                  & 700 &      &13.66 & \\
863 &  \multicolumn{6}{c}{Coadded Spectra}                  & 700 &      &13.77 & \\
906 &  6 &2007.0510 & 14:29:15.42 & 34:37:34.98 & 5 &  9000 & 260 & 1.16 &12.58 & \\
908 & 13 &2007.0219 & 14:36:24.00 & 35:27:05.00 & 3 &  5400 & 242 & 1.11 &11.99 & \\
909 &  9 &2007.0513 & 14:32:58.13 & 35:25:27.02 & 5 &  9000 & 263 & 1.14 & 6.32 & \\
910 & 10 &2007.0512 & 14:32:49.54 & 34:45:25.01 & 6 & 10800 & 267 & 1.41 & 8.90 & \\
911 & 11 &2007.0513 & 14:32:21.67 & 33:54:24.01 & 6 & 10800 & 265 & 1.33 &13.08 & \\
912 & 12 &2007.0315 & 14:33:56.00 & 33:20:45.00 & 6 & 10800 & 263 & 1.07 &13.12 & \\
914 & 14 &2007.0615 & 14:35:54.43 & 34:37:10.99 & 3 &  5400 & 246 & 1.14 &10.34 & \\
915 & 15 &2007.0424 & 14:35:37.63 & 34:04:47.98 & 5 &  9000 & 262 & 1.18 &11.72 & \\
928 &  8 &2007.0514 & 14:29:47.93 & 33:15:55.00 & 2 &  3600 & 264 & 1.02 &10.02 & \\
929 &  9 &2007.0612 & 14:32:58.13 & 35:25:27.02 & 5 &  9000 & 252 & 1.39 & 7.38 & \\
930 & 10 &2007.0616 & 14:32:49.54 & 34:45:25.01 & 6 & 10800 & 264 & 1.23 & 5.62 & \\
931 & 11 &2007.0618 & 14:32:23.69 & 33:53:23.61 & 6 & 10800 & 263 & 1.19 &11.80 & \\
934 & 14 &2007.0617 & 14:35:54.43 & 34:37:10.99 & 3 &  5400 & 263 & 1.02 &12.72 & \\
935 & 15 &2007.0614 & 14:35:40.04 & 34:04:29.98 & 5 &  9000 & 252 & 1.27 &10.93 & \\
948 &  8 &2007.0619 & 14:29:47.45 & 33:15:49.00 & 6 & 10800 & 261 & 1.18 &15.23 & \\
950 & 10 &2007.0718 & 14:32:49.54 & 34:45:25.01 & 3 &  5400 & 264 & 1.30 & 4.77 & \\
\enddata
\label{tab:passes}
\tablecomments{
For each Pass we give the closest field center from Table~\ref{tab:fields}, the RA/Dec
of the pointing, the number of exposures, and the total exposure time.  
$N_{spec}$
is the number of object spectra, Air Mass is the air mass near the middle of the
exposures, Mean SNR is the mean signal-to-noise ratio of the object spectra.}
\end{deluxetable}

\def\d{\hphantom{2}}
\begin{deluxetable}{crrcrrrrrrrr}
\rotate
\scriptsize
\tablecaption{Final Samples In 2007}
\tablewidth{0pt}
\tablehead{ Sample Name      &code06&Qshort/ &F/B/R &Total &Main &Total &Main &Total &Main &Total &Main \\
                             &      &Gshort  &      &\multicolumn{2}{c}{Sample} &\multicolumn{2}{c}{Spectra} &\multicolumn{2}{c}{Redshifts}&\multicolumn{2}{c}{Completeness}
      }
\startdata
WSRT             &      8  & 4       &                       &896   &884   &  789  &  785 &  592  &  588 &66\% &67\% \\
X-ray            &     16  & 8       &                       &3751  &3282  & 3048  & 2895 & 2424  & 2294 &65\% &70\% \\
MIPS             &     32  &16       &                       &2347  &2070  & 2125  & 1991 & 1843  & 1725 &79\% &83\% \\
IRAC             &     64  &32       &                       &5458  &4759  & 4318  & 4079 & 3174  & 2977 &58\% &63\% \\
\hline
MIPS             &    128  &1        &\d$0.3$/\d$0.5$ /30\%     &5284  &4662  & 4588  & 4484 & 4510  & 4411 &85\% &95\% \\
IRAC [8.0]          &    256  &2        &$13.8$/$13.2$/30\%     &4174  &3536  & 3645  & 3498 & 3633  & 3490 &87\% &99\%  \\
IRAC [5.8]          &    512  &4        &$15.2$/$14.7$/30\%     &4771  &4058  & 4173  & 3982 & 4110  & 3927 &88\% &98\%  \\
IRAC [4.5]          &   1024  &8        &$15.7$/$15.2$/30\%     &7261  &6215  & 6324  & 6081 & 6234  & 5999 &87\% &98\% \\
IRAC [3.6]          &   2048  &16       &$15.7$/$15.2$/30\%     &5861  &4992  & 5095  & 4882 & 4999  & 4792 &87\% &98\% \\
GALEX FUV        &   4096  &32       &$22.0$/$22.5$/30\%     & 605  & 545  &  537  &  422 &  535  &  520 &89\% &96\%   \\
GALEX NUV        &   8192  &64       &$22.0$/$21.0$/30\%     &2068  &1836  & 1838  & 1779 & 1832  & 1775 &89\% &97\%  \\
K-band           &  16384  &128      &$16.5$/$16.0$/20\%     &5676  &5399  & 5431  & 5314 & 5416  & 5302 &96\% &98\%  \\
J-band           &  32768  &256      &$18.5$/$17.5$/20\%     &4517  &4319  & 4288  & 4218 & 4278  & 4210 &95\% &98\%   \\
B-band           &  65536  &512      &$21.3$/$20.5$/20\%     &5097  &4345  & 4471  & 4278 & 4426  & 4237 &88\% &99\%   \\
R-band           & 131072  &1024     &$20.0$/$19.2$/20\%     &8904  &7480  & 7685  & 7378 & 7606  & 7304 &86\% &99\%  \\
Other I-band     & 262144  &2048     &$20.0$                 &22055 &18368 & 8428  & 8257 & 7880  & 7727 &36\% &42\%   \\
Main I-band      & 524288  &4096     &$20.0$/$18.5$/20\%     &13122 &11011 &11019  &10640 &10667  &10306 &81\% &94\%   \\
\enddata
\label{tab:samples1}
\tablecomments{ The Q/Gshort column gives the Qshort/Gshort code for the quasar (above rule) and
galaxy (below rule) samples.  For the galaxy samples, the F/B/R column gives the Faint limiting magnitude (or flux)
of the sample, the Bright magnitude limit to which it is complete, and Random sampling fraction for
the sources between the Bright and Faint magnitudes. For the 24$\mu$m galaxy sample Bright/Faint are
in mJy rather than magnitudes. The Total and Main columns give the number of targets overall and the
number inside the 15 standard sub-fields.  We list the size of each sample, the number for which spectra
were obtained, the number of successful redshift measurements and the resulting completeness. }
\end{deluxetable}

\begin{deluxetable}{llrrrrrrrrrrl}
\rotate
\scriptsize
\tablecaption{Summary of AGN Selection}
\tablewidth{0pt}
\tablehead{ Sample &Case  &Targs &Try  &Fail &Succeed &Star  &$z>0.5$ &$>1$ &$>2$ &$>3$ &$>4$ &Comment }
\startdata
WSRT        &ALL          & 896  & 789 &197  & 592   &   9 & 244 & 57 & 22  & 4 &0 &ALL \\
WSRT        &pnt          & 132  & 123 & 28  &  95   &   8 &  63 & 41 & 18  & 3 &0 &point sources \\
WSRT        &gal          & 472  & 468 &  6  & 462   &   1 & 146 &  2 &  0  & 0 &0 &bright extended\\
WSRT        &gal          & 292  & 198 &163  &  35   &   0 &  35 & 14 &  4  & 1 &0 &faint, extended \\
\hline
X-ray       &ALL          &3751  &3048 &624  &2424   & 135 &1694 &1084 &325 &57 &3  &ALL\\
X-ray       &pnt          &1907  &1685 &191  &1494   & 131 &1263 & 983 &302 &50 &3 &point sources\\
X-ray       &gal          & 848  & 751 & 10  & 741   &   4 & 256 &   7 &  3 & 2 &0 &bright, extended \\
X-ray       &gal          & 996  & 612 &423  & 189   &   0 & 175 &  94 & 20 & 5 &0 &faint, extended\\
\hline
MIPS QSO    &pnt          &2347  &2125 &282  &1843   &  41 &1353 & 871 &272 &55 &10 &point sources \\
\hline
IRAC QSO    &ALL          &5458  &4318 &1144 &3174   & 231 &2071 &1550 &526 &88 &5 &ALL\\
IRAC QSO    &1            &2887  &2571 & 398 &2173   & 207 &1573 &1294 &405 &62 &3 &point, bright red\\
IRAC QSO    &2            & 405  & 291 & 133 & 158   &   5 & 129 & 113 & 74 &13 &1 &point, faint, red\\
IRAC QSO    &3            & 691  & 429 & 153 & 276   &  17 & 146 &  62 & 26 &11 &1 &point, bright, bluer\\
IRAC QSO    &4            & 237  & 118 &   2 & 216   &   0 &  61 &   4 &  1 &0  &0 &extended, bright\\
IRAC QSO    &5            & 759  & 447 & 357 &  90   &   1 &  88 &  70 & 19 &1  &0 &extended, faint\\
\hline
ALL QSO     &             &8977  &7102 &1885 &5217   & 453 &2926 &1718 &605 &119 &13 &\\
ALL gals    &             &35177 &19447& 900 &18547  & 384 &3341 &  12 &  6 &  3 & 0 &  \\
\enddata
\label{tab:samples2}
\tablecomments{
  Quasar search yields for various samples, in some cases broken down into sub-categories.
 }
\end{deluxetable}

\begin{deluxetable}{llrrrrrrrrrrr}
\rotate
\scriptsize
\tablecaption{Summary of Selection Codes and Flags}
\tablewidth{0pt}
\tablehead{ RA &Dec  &Code06 &Code05  &Code04 &rcode &field &bstar &qirac &gal &qso &pntsrc &agngal \\}l 
\startdata
$   217.375476$ &$    32.806272$ &$  720896$ &$  720896$ &$    1536$  &$ 0$ &$ 8$ &$0$ &$0$ &$1$ &$0$ &$0$ &$1$ \\ 
$   217.893029$ &$    32.806415$ &$  724480$ &$  724480$ &$    2076$  &$ 2$ &$ 0$ &$0$ &$0$ &$1$ &$0$ &$0$ &$1$ \\ 
$   217.297011$ &$    32.806733$ &$     112$ &$     112$ &$     288$  &$ 6$ &$ 0$ &$0$ &$1$ &$0$ &$1$ &$1$ &$0$ \\ 
$   217.304623$ &$    32.806823$ &$  262144$ &$  262144$ &$    1024$  &$ 6$ &$ 0$ &$0$ &$0$ &$1$ &$0$ &$0$ &$1$ \\ 
$   216.311008$ &$    32.808259$ &$  589824$ &$  589824$ &$    1088$  &$11$ &$ 4$ &$0$ &$0$ &$1$ &$0$ &$0$ &$1$ \\ 
$   216.333753$ &$    32.806956$ &$  524288$ &$  524288$ &$      -1$  &$16$ &$ 4$ &$0$ &$0$ &$1$ &$0$ &$0$ &$1$ \\ 
$   216.393848$ &$    32.806964$ &$  724864$ &$  724864$ &$    2174$  &$12$ &$ 4$ &$0$ &$0$ &$1$ &$0$ &$0$ &$1$ \\ 
$   216.644254$ &$    32.806900$ &$  720896$ &$  720896$ &$    2112$  &$10$ &$ 4$ &$0$ &$0$ &$1$ &$0$ &$0$ &$1$ \\ 
$   217.088601$ &$    32.807106$ &$  262656$ &$  262784$ &$      -1$  &$18$ &$ 0$ &$0$ &$0$ &$1$ &$0$ &$0$ &$1$ \\ 
$   217.086404$ &$    32.807395$ &$  721152$ &$  721152$ &$    2112$  &$11$ &$ 0$ &$0$ &$0$ &$1$ &$0$ &$0$ &$1$ \\ 
$   217.479422$ &$    32.807504$ &$  524288$ &$  524288$ &$      -1$  &$ 0$ &$ 8$ &$0$ &$0$ &$1$ &$0$ &$0$ &$1$ \\ 
$   216.548230$ &$    32.807663$ &$  655360$ &$  655360$ &$    1536$  &$ 3$ &$ 4$ &$0$ &$0$ &$1$ &$0$ &$0$ &$1$ \\ 
$   217.228471$ &$    32.807886$ &$  262144$ &$  262144$ &$      -1$  &$13$ &$ 0$ &$0$ &$0$ &$1$ &$0$ &$0$ &$1$ \\ 
$   216.206669$ &$    32.808026$ &$  262144$ &$  262144$ &$      -1$  &$10$ &$ 4$ &$0$ &$0$ &$1$ &$0$ &$0$ &$1$ \\ 
$   216.199841$ &$    32.808199$ &$      16$ &$      -1$ &$      -1$  &$ 5$ &$ 4$ &$0$ &$0$ &$0$ &$1$ &$0$ &$1$ \\ 
$   216.845202$ &$    32.808131$ &$  262144$ &$  262144$ &$    1024$  &$12$ &$ 4$ &$0$ &$0$ &$1$ &$0$ &$0$ &$1$ \\ 
$   217.467032$ &$    32.808139$ &$  262272$ &$  262272$ &$    1024$  &$12$ &$ 8$ &$0$ &$0$ &$1$ &$0$ &$0$ &$1$ \\ 
$   216.328172$ &$    32.808479$ &$       0$ &$       0$ &$      -1$  &$14$ &$ 4$ &$0$ &$0$ &$1$ &$0$ &$0$ &$1$ \\ 
$   216.584225$ &$    32.808667$ &$  458752$ &$  458752$ &$    2112$  &$10$ &$ 4$ &$0$ &$0$ &$1$ &$0$ &$0$ &$1$ \\ 
$   217.618515$ &$    32.808876$ &$  524288$ &$  524288$ &$      -1$  &$ 3$ &$ 0$ &$0$ &$0$ &$1$ &$0$ &$0$ &$1$ \\ 
\enddata
\label{tab:codes}
\tablecomments{
  Code06, Code05, Code04 are the binary sample selection codes for the 2006/2007 (see \S\ref{sec:finalsample}), 2005 
  (see \S\ref{sec:middlesample}) and 2004 (see \S\ref{sec:firstsample}) survey periods, 
  rcode is the random sample code, with each index representing a randomly selected 5\% of the
  sources, field is the sub-field number, where 0 indicates that it is outside the standard
  fields.  ``qirac'' gives the IRAC AGN selection sub-code for the 2006/2007 season.  The codes
  gal, qso, pntsrc and agngal indicate whether the source was counted as a galaxy target,
  an AGN target, a point source, or a faint ($\hbox{I}>20$~mag) extended AGN candidate.  
  See \S\ref{sec:design} for a detailed definitions and discussions of the entries.
  The RA and Dec in this and later tables are the NDWFS DR3 I-band coordinates.
  The on-line version contains the complete table.
 }
\end{deluxetable}

\begin{deluxetable}{rrrrrrrrrrrrrrr}
\scriptsize
\tablecaption{Magnitudes And Fluxes}
\rotate
\scriptsize
\tablewidth{0pt}
\tablehead{ X &FUV &NUV &B$_W$ &R &I &z' &J &K1 &K2 &[3.6] &[4.5] &[5.8] &[8.0] &24}
\startdata
$  -10$ &$-10.00$ &$-10.00$ &$ 20.71$ &$ 19.29$ &$ 18.77$ &$ 19.13$ &$-10.00$ &$-10.00$ &$-10.00$ &$ 16.63$ &$ 16.45$ &$-10.00$ &$ 14.02$ &$-10.000$ \\ 
$  -10$ &$-10.00$ &$ 24.52$ &$ 20.78$ &$ 18.15$ &$ 17.43$ &$ 17.63$ &$-10.00$ &$-10.00$ &$-10.00$ &$ 14.53$ &$ 14.46$ &$ 14.40$ &$ 14.27$ &$-10.000$ \\ 
$    3$ &$-10.00$ &$-10.00$ &$ 18.52$ &$ 18.20$ &$ 17.87$ &$ 18.38$ &$-10.00$ &$-10.00$ &$-10.00$ &$ 14.66$ &$ 13.72$ &$ 12.85$ &$ 11.81$ &$ 3.474$ \\ 
$  -10$ &$-10.00$ &$-10.00$ &$ 20.89$ &$ 19.77$ &$ 19.31$ &$ 19.71$ &$-10.00$ &$-10.00$ &$-10.00$ &$ 17.44$ &$ 17.51$ &$-10.00$ &$-10.00$ &$-10.000$ \\ 
$  -10$ &$-10.00$ &$-10.00$ &$ 20.43$ &$ 19.71$ &$ 17.27$ &$ 19.77$ &$-10.00$ &$-10.00$ &$-10.00$ &$-10.00$ &$ 16.81$ &$-10.00$ &$ 15.47$ &$-10.000$ \\ 
$  -10$ &$-10.00$ &$-10.00$ &$-10.00$ &$ 20.94$ &$ 18.22$ &$ 21.51$ &$-10.00$ &$-10.00$ &$-10.00$ &$-10.00$ &$-10.00$ &$-10.00$ &$-10.00$ &$-10.000$ \\ 
$  -10$ &$-10.00$ &$-10.00$ &$ 18.98$ &$ 17.32$ &$ 16.64$ &$ 16.93$ &$-10.00$ &$-10.00$ &$-10.00$ &$ 14.00$ &$ 13.94$ &$ 13.29$ &$ 10.82$ &$ 2.775$ \\ 
$  -10$ &$-10.00$ &$-10.00$ &$ 20.45$ &$ 18.64$ &$ 17.89$ &$-10.00$ &$-10.00$ &$-10.00$ &$-10.00$ &$ 16.53$ &$ 16.52$ &$-10.00$ &$ 14.93$ &$-10.000$ \\ 
$  -10$ &$-10.00$ &$-10.00$ &$ 22.25$ &$ 20.01$ &$ 19.15$ &$ 19.29$ &$-10.00$ &$-10.00$ &$-10.00$ &$ 15.58$ &$ 15.45$ &$ 14.27$ &$ 14.13$ &$ 0.467$ \\ 
$  -10$ &$-10.00$ &$-10.00$ &$ 20.29$ &$ 18.59$ &$ 17.95$ &$ 18.14$ &$-10.00$ &$-10.00$ &$-10.00$ &$ 15.55$ &$ 15.25$ &$-10.00$ &$ 12.94$ &$-10.000$ \\ 
$  -10$ &$-10.00$ &$ 23.07$ &$ 22.25$ &$ 20.52$ &$ 19.88$ &$ 20.32$ &$-10.00$ &$-10.00$ &$-10.00$ &$ 17.18$ &$ 16.51$ &$-10.00$ &$ 15.00$ &$-10.000$ \\ 
$  -10$ &$-10.00$ &$-10.00$ &$ 22.08$ &$ 19.92$ &$ 19.21$ &$ 19.38$ &$-10.00$ &$-10.00$ &$-10.00$ &$ 16.03$ &$ 15.72$ &$ 15.79$ &$ 15.23$ &$-10.000$ \\ 
$  -10$ &$-10.00$ &$-10.00$ &$ 23.11$ &$ 20.25$ &$ 19.34$ &$ 19.54$ &$-10.00$ &$-10.00$ &$-10.00$ &$ 16.14$ &$ 16.05$ &$ 16.14$ &$-10.00$ &$-10.000$ \\ 
$  -10$ &$-10.00$ &$-10.00$ &$ 23.73$ &$ 20.52$ &$ 19.40$ &$ 19.83$ &$-10.00$ &$-10.00$ &$-10.00$ &$-10.00$ &$ 15.99$ &$-10.00$ &$ 15.74$ &$-10.000$ \\ 
$    4$ &$-10.00$ &$-10.00$ &$ 22.31$ &$ 22.04$ &$ 21.53$ &$ 21.95$ &$-10.00$ &$-10.00$ &$-10.00$ &$-10.00$ &$ 16.95$ &$-10.00$ &$ 15.44$ &$-10.000$ \\ 
$  -10$ &$-10.00$ &$-10.00$ &$ 21.50$ &$ 19.65$ &$ 18.98$ &$ 19.35$ &$-10.00$ &$-10.00$ &$-10.00$ &$ 16.39$ &$ 16.21$ &$ 16.20$ &$ 13.98$ &$ 0.342$ \\ 
$  -10$ &$-10.00$ &$ 23.29$ &$ 21.77$ &$ 19.82$ &$ 19.10$ &$ 19.39$ &$-10.00$ &$-10.00$ &$-10.00$ &$ 15.87$ &$ 15.46$ &$ 15.15$ &$ 13.67$ &$ 0.693$ \\ 
$  -10$ &$-10.00$ &$-10.00$ &$ 23.78$ &$ 14.81$ &$ 13.76$ &$ 20.58$ &$-10.00$ &$-10.00$ &$-10.00$ &$ 13.68$ &$ 13.39$ &$-10.00$ &$-10.00$ &$-10.000$ \\ 
$  -10$ &$-10.00$ &$-10.00$ &$ 20.36$ &$ 19.15$ &$ 18.67$ &$ 19.07$ &$-10.00$ &$-10.00$ &$-10.00$ &$ 16.81$ &$ 16.56$ &$-10.00$ &$ 14.24$ &$-10.000$ \\ 
$  -10$ &$-10.00$ &$-10.00$ &$ 23.09$ &$ 20.57$ &$ 19.79$ &$ 19.84$ &$-10.00$ &$-10.00$ &$-10.00$ &$ 16.77$ &$ 16.57$ &$-10.00$ &$ 14.78$ &$-10.000$ \\ 
\enddata
\label{tab:mags}
\tablecomments{
 All entries except the X-ray column (X, counts) and $24\mu$m column (24, mJy) are in
 Vega magnitudes. An entry of $-10$ means no data or below the magnitude limit of the
 survey.  X is from XBo\"otes (\citealt{Murray2005}, \citealt{Kenter2005},
 \citealt{Brand2006}), FUV and NUV are from GALEX (\citealt{Martin2005}), B$_W$, R, I and K2 are 
 from the NDWFS DR3 (\citealt{Jannuzi1999}), z' is from zBo\"otes (\citealt{Cool2007}),
 J and K1 are from FLAMEX (\citealt{Elston2006}), [3.6]-[8.0] are from the IRAC
 Shallow Survey (\citealt{Eisenhardt2004}) and the MIPS $24\mu$m fluxes are
 from \cite{Soifer2004}.  The on-line version contains the complete table.
 }
\end{deluxetable}

\begin{deluxetable}{llcrrrrrrrrrrrr}
\rotate
\scriptsize
\tablecaption{Redshifts}
\tablewidth{0pt}
\tablehead{ RA &Dec  &N &\multicolumn{4}{c}{Spectrum 1} 
  &\multicolumn{4}{c}{Spectrum 2} 
  &\multicolumn{4}{c}{Spectrum 3}  \\
 & & &$z$ &$S/N$ &pass &ap &$z$ &$S/N$ &pass &ap &$z$ &$S/N$ &pass &ap }
\startdata
$   217.375476$ &$    32.806272$ &$ 1$ &$   0.219$ &$   9.009$ &$308$ &$230$ \\ 
$   217.893029$ &$    32.806415$ &$ 0$ \\ 
$   217.297011$ &$    32.806733$ &$ 1$ &$   1.257$ &$  34.483$ &$108$ &$100$ \\ 
$   217.304623$ &$    32.806823$ &$ 1$ &$   0.136$ &$  21.703$ &$948$ &$212$ \\ 
$   216.311008$ &$    32.808259$ &$ 2$ &$   0.352$ &$   2.745$ &$604$ &$100$ &$  -1.352$ &$   2.806$ &$423$ &$225$ \\ 
$   216.333753$ &$    32.806956$ &$ 0$ \\ 
$   216.393848$ &$    32.806964$ &$ 2$ &$   0.094$ &$  20.408$ &$304$ &$232$ &$   0.095$ &$   0.000$ &$  0$ &$  1$ \\ 
$   216.644254$ &$    32.806900$ &$ 1$ &$   0.131$ &$   3.534$ &$204$ &$ 90$ \\ 
$   217.088601$ &$    32.807106$ &$ 0$ \\ 
$   217.086404$ &$    32.807395$ &$ 1$ &$   0.132$ &$  15.873$ &$304$ &$177$ \\ 
$   217.479422$ &$    32.807504$ &$ 1$ &$   0.346$ &$   5.782$ &$417$ &$ 56$ \\ 
$   216.548230$ &$    32.807663$ &$ 1$ &$   0.248$ &$   9.615$ &$304$ &$213$ \\ 
$   217.228471$ &$    32.807886$ &$ 0$ \\ 
$   216.206669$ &$    32.808026$ &$ 1$ &$   0.534$ &$   2.297$ &$604$ &$102$ \\ 
$   216.199841$ &$    32.808199$ &$ 3$ &$   1.954$ &$   5.503$ &$862$ &$272$ &$   1.954$ &$   4.998$ &$824$ &$250$ &$   1.954$ &$   1.768$ &$804$ &$211$ \\ 
$   216.845202$ &$    32.808131$ &$ 2$ &$   0.217$ &$  19.907$ &$862$ &$231$ &$   0.217$ &$  16.113$ &$804$ &$165$ \\ 
$   217.467032$ &$    32.808139$ &$ 1$ &$   0.346$ &$   9.655$ &$608$ &$181$ \\ 
$   216.328172$ &$    32.808479$ &$ 1$ &$  -4.218$ &$   3.240$ &$404$ &$199$ \\ 
$   216.584225$ &$    32.808667$ &$ 1$ &$   0.170$ &$   9.615$ &$304$ &$220$ \\ 
$   217.618515$ &$    32.808876$ &$ 0$ \\ 
\enddata
\label{tab:redshift}
\tablecomments{
 $N$ indicates the number of spectra taken.  The redshift $z$, signal-to-noise ratio $S/N$, 
 pass and aperture codes (see \S\ref{sec:data}) 
 are then given for up to the first three spectra in order of decreasing
 $S/N$.  Pipeline redshifts that did not pass visual inspection are reported 
 as $-1-z$, where $z$ was the pipeline redshift estimate. 
  The on-line version contains the complete table. 
 }
\end{deluxetable}

\begin{deluxetable}{llrrrrrrrr}
\scriptsize
\tablecaption{Photometric Redshifts}
\tablewidth{0pt}
\tablehead{ RA &Dec  &$N_{band}$ &$z_{phot}$ &$\chi^2$ &E &Sbc &Im &AGN &$E(B-V)$ \\
               &     &           &           &         &\multicolumn{4}{c}{Luminosity $L/10^{10}L_\odot$} &mag 
 }
\startdata
$   217.375476$ &$    32.806272$ &$ 9$ &$ 0.160$ &$ 1.443$ &$ 0.238$ &$ 0.168$ &$ 0.351$ &$ 0.052$ &$ 0.184$ \\ 
$   217.893029$ &$    32.806415$ &$ 9$ &$ 0.260$ &$ 2.347$ &$ 6.906$ &$ 0.000$ &$ 0.000$ &$ 0.085$ &$ 0.359$ \\ 
$   217.297011$ &$    32.806733$ &$10$ &$ 0.120$ &$15.820$ &$ 0.000$ &$ 0.000$ &$ 1.090$ &$ 0.039$ &$ 0.048$ \\ 
$   217.304623$ &$    32.806823$ &$ 9$ &$ 0.220$ &$ 9.626$ &$ 0.242$ &$ 0.000$ &$ 0.715$ &$ 0.072$ &$ 0.000$ \\ 
$   216.311008$ &$    32.808259$ &$ 9$ &$ 0.580$ &$49.400$ &$ 0.000$ &$ 1.348$ &$ 1.464$ &$ 0.190$ &$ 0.184$ \\ 
$   216.333753$ &$    32.806956$ &$ 9$ &$ 0.600$ &$22.913$ &$ 1.230$ &$ 0.000$ &$ 0.541$ &$ 0.197$ &$ 0.147$ \\ 
$   216.393848$ &$    32.806964$ &$ 9$ &$ 0.140$ &$13.895$ &$ 0.000$ &$ 3.627$ &$ 0.000$ &$ 0.046$ &$ 0.450$ \\ 
$   216.644254$ &$    32.806900$ &$ 7$ &$ 0.220$ &$ 0.990$ &$ 0.535$ &$ 0.055$ &$ 0.336$ &$ 0.072$ &$ 0.000$ \\ 
$   217.088601$ &$    32.807106$ &$ 8$ &$ 0.460$ &$ 0.887$ &$ 4.222$ &$ 2.537$ &$ 0.000$ &$ 0.151$ &$ 0.094$ \\ 
$   217.086404$ &$    32.807395$ &$ 8$ &$ 0.140$ &$ 1.463$ &$ 0.663$ &$ 0.337$ &$ 0.220$ &$ 0.046$ &$ 0.000$ \\ 
$   217.479422$ &$    32.807504$ &$ 9$ &$ 0.280$ &$ 2.062$ &$ 0.024$ &$ 1.042$ &$ 0.270$ &$ 0.092$ &$ 0.000$ \\ 
$   216.548230$ &$    32.807663$ &$ 8$ &$ 0.340$ &$ 0.408$ &$ 2.185$ &$ 0.435$ &$ 0.327$ &$ 0.112$ &$ 0.000$ \\ 
$   217.228471$ &$    32.807886$ &$ 8$ &$ 0.440$ &$ 0.507$ &$ 4.339$ &$ 0.000$ &$ 0.001$ &$ 0.144$ &$ 0.094$ \\ 
$   216.206669$ &$    32.808026$ &$ 8$ &$ 0.500$ &$ 1.835$ &$ 4.987$ &$ 0.000$ &$ 0.000$ &$ 0.164$ &$ 0.230$ \\ 
$   216.199841$ &$    32.808199$ &$ 8$ &$ 0.680$ &$ 2.961$ &$ 0.000$ &$ 1.143$ &$ 0.000$ &$ 0.223$ &$ 0.048$ \\ 
$   216.845202$ &$    32.808131$ &$ 9$ &$ 0.220$ &$ 1.132$ &$ 0.600$ &$ 0.702$ &$ 0.166$ &$ 0.072$ &$ 0.000$ \\ 
$   217.467032$ &$    32.808139$ &$ 9$ &$ 0.280$ &$ 0.386$ &$ 0.034$ &$ 2.940$ &$ 0.000$ &$ 0.092$ &$ 0.147$ \\ 
$   216.328172$ &$    32.808479$ &$ 7$ &$ 1.000$ &$86.301$ &$ 2.305$ &$12.494$ &$ 0.000$ &$ 0.328$ &$ 0.000$ \\ 
$   216.584225$ &$    32.808667$ &$10$ &$ 0.100$ &$ 4.412$ &$ 0.105$ &$ 0.039$ &$ 0.167$ &$ 0.033$ &$ 0.184$ \\ 
$   217.618515$ &$    32.808876$ &$ 9$ &$ 0.360$ &$ 2.089$ &$ 1.306$ &$ 0.870$ &$ 0.070$ &$ 0.118$ &$ 0.000$ \\ 
\enddata
\label{tab:photoz}
\tablecomments{
 Photometric redshift and SED decompositions following \cite{Assef2010}.
 $N_{band}$ is the number of bands used to derive the photometric redshift $z_{phot}$ and $\chi^2$
 is the goodness of fit at the photometric redshift.  The E, Sbc, Im and AGN columns give the
 contributions of these templates to the SED in units of $10^{10}L_\odot$.  By definition the
 contributions are always $\geq 0$, and they are calculated for the spectroscopic redshift 
 if it is known.  The AGN luminosity is only calculated redward
 of the Lyman limit.  $E(B-V)$ is the extinction applied to the AGN template.  
 The on-line version contains the complete table.
 }
\end{deluxetable}

\begin{deluxetable}{rrrrr}
\scriptsize
\tablecaption{Spectra}
\tablewidth{0pt}
\tablehead{ Pass &Aperture  &$\lambda (\AA) $ &$(\hbox{Error})^{-1}$  &Spectrum }
\startdata
\multicolumn{5}{c}{Only in Electronic Version}\\
\enddata
\label{tab:spectra}
\tablecomments{
 The spectra as available.  These are a mixture of results from the two pipelines
 and fluxed and un-fluxed spectra.  Depending on the pipeline, the Error entry is
 either the inverse of the estimated noise in the spectrum or a masking flag.  
 Fully homogenizing these reductions is beyond the scope of the present paper.
 }
\end{deluxetable}

\begin{deluxetable}{rrcccrr}
\scriptsize
\tablecaption{Completeness and K-Corrections for the Galaxy Samples}
\tablewidth{0pt}
\tablehead{ RA &Dec  &\multicolumn{3}{c}{Completeness Corrections} &$z_{max}$ &$V_{max}$ \\
               &     &Spec        &Sparse      &Fiber           &          &$10^6 h^{-3}$~Mpc$^3$ 
 }
\startdata
$   217.375476$ &$    32.806272$ &$ 1.029$ &$ 5.000$ &$ 1.004$ &$ 0.391$ &$ 0.928$ \\ 
$   217.893029$ &$    32.806415$  \\ 
$   217.297011$ &$    32.806733$  \\ 
$   217.304623$ &$    32.806823$  \\ 
$   216.311008$ &$    32.808259$ &$ 1.054$ &$ 1.000$ &$ 1.004$ &$ 2.346$ &$47.569$ \\ 
$   216.333753$ &$    32.806956$  \\ 
$   216.393848$ &$    32.806964$ &$ 1.000$ &$ 1.000$ &$ 1.002$ &$ 0.423$ &$ 1.148$ \\ 
$   216.644254$ &$    32.806900$ &$ 1.000$ &$ 1.000$ &$ 1.004$ &$ 0.335$ &$ 0.611$ \\ 
$   217.088601$ &$    32.807106$  \\ 
$   217.086404$ &$    32.807395$ &$ 1.000$ &$ 1.000$ &$ 1.000$ &$ 0.335$ &$ 0.609$ \\ 
$   217.479422$ &$    32.807504$ &$ 1.040$ &$ 5.000$ &$ 1.307$ &$ 0.364$ &$ 0.768$ \\ 
$   216.548230$ &$    32.807663$ &$ 1.003$ &$ 5.000$ &$ 1.004$ &$ 0.355$ &$ 0.719$ \\ 
$   217.228471$ &$    32.807886$  \\ 
$   216.206669$ &$    32.808026$ &$ 1.000$ &$ 0.000$ &$ 1.000$ &$ 0.687$ &$ 3.975$ \\ 
$   216.199841$ &$    32.808199$  \\ 
$   216.845202$ &$    32.808131$  \\ 
$   217.467032$ &$    32.808139$ &$ 1.006$ &$ 1.000$ &$ 1.307$ &$ 0.521$ &$ 1.987$ \\ 
$   216.328172$ &$    32.808479$  \\ 
$   216.584225$ &$    32.808667$ &$ 1.006$ &$ 1.000$ &$ 1.002$ &$ 0.322$ &$ 0.547$ \\ 
$   217.618515$ &$    32.808876$  \\ 
\enddata
\label{tab:stats}
\tablecomments{
 Factors needed to properly weight the galaxy samples following \cite{Cool2011}. 
 There are three completeness corrections factors.  ``Spec'' corrects for the failure to determine a
 redshift when a spectrum was obtained, ``Sparse'' corrects for the target sparse sampling 
 weighting, and ``Fiber'' corrects for how the local target density affects the assignment
 of fibers.  The maximum redshift at which the source would have included the target
 is $z_{max}$ and this corresponds to a volume of $V_{max}$ in units of $10^6 h^{-3}$~Mpc$^3$
 for an $\Omega=0.3$, $\Lambda=0.7$ cosmological model and the survey area of $7.60$~deg$^2$
 used by \cite{Cool2011}.  These are calculated using {\bf kcorrect v4$\_2$} 
 (\citealt{Blanton2007}) based on the B$_w$RI photometry.
 }
\end{deluxetable}

\end{document}